\documentclass[printer,dvipsnames]{aa}
\usepackage{xcolor}
\usepackage{mathtools}
\usepackage{ulem}
\usepackage{natbib}
\usepackage{color}
\usepackage{graphicx}
\usepackage{txfonts}
\newcommand{\gaia}{\textit{Gaia}}

\begin{document}

\title{{\it Gaia} DR2 unravels incompleteness of nearby cluster population:\\New open clusters in the direction of Perseus}

   \author{T. Cantat-Gaudin\inst{\ref{UB}},
       A. Krone-Martins\inst{\ref{Lisbon}},
       N. Sedaghat\inst{\ref{Freiburg}},
       A. Farahi\inst{\ref{CMU}},
       R. S. de Souza\inst{\ref{UNC}},
       R. Skalidis\inst{\ref{Greece1}}$^{,}$\inst{\ref{Greece2}},
       A. I. Malz\inst{\ref{NYU1}}$^{,}$\inst{\ref{NYU2}},\\
       S. Mac\^edo\inst{\ref{IFPE}},
       B. Moews\inst{\ref{Edinburgh}},
       C. Jordi\inst{\ref{UB}}, 
       A. Moitinho\inst{\ref{Lisbon}},
       A. Castro-Ginard\inst{\ref{UB}},
       E. E. O. Ishida\inst{\ref{Auvergne}},
       C. Heneka\inst{\ref{Pisa}}, \\
       A. Boucaud\inst{\ref{Paris1}}$^{,}$\inst{\ref{Paris2}},
       A. M. M. Trindade\inst{\ref{PT2}},
       for the COIN collaboration
        }

  \institute{
		Institut de Ci\`encies del Cosmos, Universitat de Barcelona (IEEC-UB), Mart\'i i Franqu\`es 1, E-08028 Barcelona, Spain\label{UB}
		\\
		 \email{tcantat@fqa.ub.edu}
		 \and
  		 CENTRA, Faculdade de Ci\^encias, Universidade de Lisboa, Ed. C8, Campo Grande, 1749-016 Lisboa, Portugal\label{Lisbon}
  		 \and
  		 Department of Computer Science, University of Freiburg, Georges-Koehler-Allee 052, 79110 Freiburg, Germany\label{Freiburg}
  		 \and
  		 McWilliams Center for Cosmology, Department of Physics, Carnegie Mellon University,  5000 Forbes Ave., Pittsburgh, PA 15213, USA\label{CMU}
  		 \and
  		 Department of Physics \& Astronomy, University of North Carolina at Chapel Hill, NC 27599-3255, USA\label{UNC}
  		  \and
  		 Department of Physics and Institute for Theoretical and Computational Physics, University of Crete, GR-71003 Heraklion Greece\label{Greece1}
  		 \and 
  		 Foundation for Research and Technology – Hellas, IESL, Voutes, GR-7110 Heraklion, Greece\label{Greece2}
  		 \and 
  		 Center for Cosmology and Particle Physics, New York University, 726 Broadway, New York, NY 10004, USA \label{NYU1}
  		 \and 
  		 Department of Physics, New York University, 726 Broadway, New York, NY 10004, USA \label{NYU2}
		\and
		 Instituto Federal de Educa\c{c}\~ao, Ci\^encia e Tecnologia de Pernambuco, IFPE, 50670-430, Recife-PE, Brazil\label{IFPE}
  		 \and
  		 Institute for Astronomy, University of Edinburgh, Blackford Hill, Edinburgh EH9 3HJ, UK\label{Edinburgh}
  		 \and 
  		 Universit\'e Clermont Auvergne, CNRS/IN2P3, LPC, F-63000 Clermont-Ferrand, France\label{Auvergne}
  		 \and
		Scuola Normale Superiore, Piazza dei Cavalieri 7, 56126 Pisa, Italy\label{Pisa}
  		 \and
		LAL, Univ. Paris-Sud, CNRS/IN2P3, Universit{\'e} Paris-Saclay, Orsay, France\label{Paris1}
		\and
		 Paris-Saclay Center for Data Science, Universit{\'e} Paris-Saclay, Orsay, France \label{Paris2}
  		 \and
		 Instituto de Astrofísica e Ciências do Espaço, Universidade do Porto, CAUP, Rua das Estrelas, 4150-762 Porto, Portugal \label{PT2}
  		 }

   \date{Received October 12, 2018; accepted ???, ???}

  \abstract{Open clusters (OCs) are popular tracers of the structure and evolutionary history of the Galactic disk. The OC population is often considered to be  complete within 1.8 kpc of the Sun. The recent \textit{Gaia} Data Release 2 (DR2) allows the latter claim to be challenged.}
   {We perform a systematic search for new OCs in the direction of Perseus using precise and accurate astrometry from {\it Gaia} DR2.}
   {We implement a coarse-to-fine search method. 
   First, we exploit spatial proximity using a fast density-aware partitioning of the sky via a $k$-d tree in the spatial domain of Galactic coordinates, $(l,b)$. 
   Secondly, we employ a Gaussian mixture model in the proper motion space to quickly tag  fields around OC candidates. 
   Thirdly, we apply an unsupervised membership assignment method, UPMASK, to scrutinise the candidates. 
   We visually inspect colour-magnitude diagrams to validate the detected objects. Finally, we perform a diagnostic to quantify the significance of each identified overdensity
   in proper motion and in parallax space.}
   {We report the discovery of 41 new stellar clusters. This represents an increment of at least 20\% of the previously known OC population in this volume of the Milky Way. 
   We also report on the clear identification of NGC 886, an object previously considered an asterism. 
   This study challenges the previous claim of a near-complete sample of open clusters up to 1.8 kpc. 
   Our results reveal that this claim requires revision, and a complete census of nearby open clusters is yet to be found. }
   {}
\keywords{open clusters - stars: solar neighbourhood, methods:data analysis, statistical–techniques}
\titlerunning{New open clusters in the direction of Perseus}
\authorrunning{T. Cantat-Gaudin et al.}
\maketitle
\section{Introduction}

Galactic stellar clusters, traditionally called open clusters (OCs), are  tracers of the structure and evolution of the Milky Way \citep[e.g.][]{Janes82,2005ApJ...629..825D,Piskunov06,Moitinho10}, playing a fundamental role in studies of star formation environment and evolution \citep[e.g.][]{2010RSPTA.368..713L}. 
They are found primarily in the Galactic plane and are composed of dozens to several thousands of stars of similar age, metallicity, kinematics, and distance. 
As such, they provide a testbed for  stellar evolution models.

It has been often stated in the literature that the census of the OC population in the solar neighbourhood is complete. \citet{Piskunov06} put the completenes radius at 850\,pc. This has been contested by \citet{Moitinho10}, who shows that, based on the photometry used by \citet{Piskunov06}, many sparse and/or old open clusters are expected to be missed. \citet{Kharchenko2013}, working with 2MASS photometry \citep{Skrutskie06}, made the claim that the OC sample is almost complete within 1.8\,kpc from the Sun. Few discoveries of nearby OCs had been made since the publications of \citet{Alessi2003} and \citet{Kharchenko2005} \citep[with the notable exception of][who identified nine new objects within 500\,pc]{Roeser16}, making the completeness claim appear plausible. While the astrometric and photometric analysis followed in the aforementioned works provide a well-established methodology for identifying members and deriving properties of clusters, a significant part of the problem seems to arise from the data used in those studies. The catalogues they are based on, such as Hipparcos \citep{ESA97}, 2MASS \citep{Skrutskie06}, PPMXL \citep{Roeser10}, or UCAC \citep[e.g.][]{Zacharias13} have brought an enormous contributions to astronomy, but as any data set they have sensitivity limitations and should not be overinterpreted. As noted in \citet{Moitinho10}, many objects will fall on the borderline of the data sensitivity limits, which causes incompleteness and creates false positives in the sample.

The improvement in quality and the depth of the recently published \gaia\ Data Release 2 \citep[DR2,][]{2018A&A...616A...1G} allow us to identify new objects and address the question of the reality of previously identified objects. In a recent publication, \citet{Castro18} reported on the discovery of 11 OCs closer than 500 pc in the \gaia\ DR1 data \citep{2016A&A...595A...2G}. \citet{CantatGaudin18gdr2} have identified 34 new OCs within 2\,kpc in the \gaia\ DR2 data, and shown that the samples on which the completeness analyses are based are also highly contanimated by false positives.

In this study we search for known and unknown stellar clusters making use of stars in the \gaia\ DR2 catalogue up to magnitude $G=18.0$, using a coarse-to-fine search methodology combining machine learning and statistical techniques. 
Recent data releases of \gaia\ have changed the state of the field by providing accurate and precise measurements of stellar population kinematics and parallaxes. 
We use the spatial, kinematic, and parallax information in a hierarchical and automated fashion to scan the sky, flag the candidates, and validate them. 
To validate or reject the flagged candidates, we employ the UPMASK method \citep{2014A&A...561A..57K}, which uses all of the above information simultaneously. 
We neither make use of photometric data to detect clusters nor to perform stellar membership. 
The photometric information is employed solely for an independent validation of each OC candidate by visual inspection of their colour-magnitude diagrams.

In Sect.~\ref{Data}, we briefly present the \gaia\ DR2 data and the source selection criteria adopted in this study. 
The methods developed and employed to search for OCs are detailed in Sect.~\ref{Method}. 
We present the results and comment on some specific objects in Sect.~\ref{Discussion}. 
Finally, we summarise our findings in Sect.~\ref{Conclusions}. 
The information of all discovered objects can be consulted in the appendix, or retrieved in electronic form, including through Virtual Observatory services, from  CDS\footnote{\href{ftp://cdsarc.u-strasbg.fr/XXX}{ftp://cdsarc.u-strasbg.fr/XXX} or via \href{http://cdsweb.ustrasbg.fr/cgi-bin/gcat?J/A+A/XXX}{http://cdsweb.ustrasbg.fr/cgi-bin/gcat?J/A+A/XXX}.}.

\section{Data}\label{Data}

The ESA \gaia\ space mission \citep{2016A&A...595A...1G} is producing an unprecedented all-sky survey in terms of its sheer size, dimensionality and history-changing astrometric precision and accuracy. 
In particular, the most recent Data Release, \gaia\ DR2, contains more than 1.6 billion sources as faint as $G\sim21$, providing five-parameter astrometric solutions accurate to hundreds of micro-arcseconds, as well as magnitudes in three photometric bands, for more than 1.3 billion sources. \citep{2018A&A...616A...1G}. 
This unprecedented precision facilitates the identification of stellar clusters by increasing the contrast between the cluster members and the field objects in the proper motion and parallax space. 

The \gaia\ DR2 data were retrieved in two different ways. 
First, we used the \gaia\ Archive bulk retrieval data facility\footnote{\href{http://cdn.gea.esac.esa.int/Gaia/gdr2/gaia\_source/csv/}{http://cdn.gea.esac.esa.int/Gaia/gdr2/gaia\_source/csv/}} to obtain the entire region defined by galactic coordinates $l \in [120^\circ; 200^\circ]$ and $b \in [-10^\circ; 10^\circ]$, and magnitudes $G\le18$. This magnitude cut corresponds to typical astrometric uncertainties better than 0.3\,mas\,yr$^{-1}$ in proper motion, and 0.15\,mas in parallax. 
From these data, we selected the information for spatial tiling and candidate flagging (positions $l, b$, proper motions $\mu_{\alpha*}, \mu_{\delta}$, and parallax, $\varpi$; see Sects.~\ref{tiling} and \ref{cflagging}). 
Secondly (described in Sect.~\ref{canalysis}), we validated the candidates querying data through the \gaia\ archive facility at ESAC \citep{2017A&C....21...22S} with {\tt pygacs}\footnote{\href{https://github.com/Johannes-Sahlmann/pygacs}{https://github.com/Johannes-Sahlmann/pygacs}} to extract the positions ($\alpha, \delta, l, b$), parallaxes ($\varpi$), proper motions ($\mu_{\alpha*}, \mu_\delta$), fluxes in the $G$, $G_{BP}$ and $G_{RP}$ passbands, and their associated uncertainties and covariances, from the table {\tt gaiadr2.gaia\_source} via ADQL \citep{ADQL} queries launched using the TAP protocol \citep{TAP}. 

\section{Method}\label{Method}

To perform a systematic search for OCs, we design and employ a coarse-to-fine search algorithm, tailored to be used on astrometric catalogues. 
Our algorithm consists of three steps: i) fast density-aware sky tiling to exploit the proximity of cluster members on the plane of the sky; ii) OC candidate region flagging based on proper motion similarity via a fast recommendation system; iii) cluster membership assignment and independent validation. 
We detail these steps in the following subsections.

\subsection{Density-aware spatial tiling}\label{tiling}

As the very first step, we  partition the sample into computationally tractable subsets, i.e. tiles. 
Partitioning datasets of \gaia\ size requires an extremely fast and scalable method. 
At the same time, we would like to avoid loss of spatial information, or, informally speaking, splitting a potential cluster across many tiles, which lowers the signal-to-noise ratio. 
Hence, the adopted algorithm should be aware of the spatial spread of data points. 
To this end,  we employ $k$-d trees for spatial partitioning \citep{Bentley1975}.

$k$-d trees are data structures based on binary trees \citep{garnier2009discrete}, traditionally used for fast nearest neighbour searches. 
In this work, however, we use the accompanying fast partitioning algorithm to create a rectangular tessellation of the target sky region, which preserves a predefined minimum number of stars in each tile.
$k$-d trees are known to perform best when the number of dimensions ($k$) is not large \citep{friedman1977algorithm,sproull1991refinements}. 
We use the two galactic spatial coordinates $l$ and $b$, hence $k = 2$ to partition the region of interest with $\sim15$ million sources. 
The algorithm outputs 2048 tiles, each containing $\sim7400$ stars on average.\footnote{We make use of the \textit{SciPy} \citep{scipy} implementation of the $k$-d tree \citep[cKDTree algorithm from][]{maneewongvatana1999s}.}

\subsection{Candidate flagging}\label{cflagging}

To quickly search over 2048 tiles and flag open cluster candidates, we employ a Gaussian mixture model \citep[GMM; e.g.][]{Pearson71,Melchior:2016GMM} in the proper motion space $(\mu_{\alpha*},\mu_{\delta})$. 
Here, the GMM is used to decrease the clutter and information overload, rather than as an optimal model of the proper motion distribution. The Gaussian covariances are used as the score metric to decide on the follow up analysis of the field.

A GMM is a parametric model that consists of a weighted sum of Gaussian components. 
The GMM allows us to separate a candidate region from the background and foreground sources because the measured stellar proper motions of OC members can be roughly approximated by a Gaussian distribution, with means equalling the bulk proper motion of the OC, and with a variance smaller than that of the field stars. 
Therefore, GMMs are a natural choice to flag candidate regions in which an OC may exist. 
This technique has been traditionally used to study and search OCs  \citep{1958AJ.....63..387V,1971A&A....14..226S}, and has been continuously applied in astronomy \citep[e.g.][]{2010A&A...516A...3K,2017A&A...597A..90D,deSouza2017}.

We adopt GMMs with 10 multivariate Gaussian components\footnote{The results are insensitive to the exact number of Gaussians, as long as it is large enough to disentangle the background.} to perform one global search in the entire data within each tile, and multiple targeted searches in parallax bins of 0.2\,mas (increased by steps of 0.2\,mas to contain at least 1000 objects per bin). 
The targeted searches mitigate false negatives caused by the relative signal-to-noise ratio of the candidate regions when compared to the field population. 
Since we expect members of the same stellar cluster to share similar parallaxes, this parallax screening increases the search efficiency. 
Finally, the candidate region is kept for further scrutiny if its Gaussian component exhibits a variance smaller than 0.1 mas$^{2}$\,yr$^{-2}$ in both $\mu_{\alpha}*$ and $\mu_{\delta}$ (corresponding to a standard deviation of $\sim0.3$ mas\,yr$^{-1}$), and if it has at least ten members falling within its interquartile range.

\subsection{Candidate analysis}\label{canalysis}

To further analyse each candidate region, we employ an unsupervised method that relies on minimal physical assumptions about stellar clusters called UPMASK \citep{2014A&A...561A..57K}. 
The key assumptions are that the cluster member stars must share some common properties, thereby being clustered in some parameter spaces (here proper motion and parallax), and that the spatial distribution of the member stars should be incompatible with a uniform spatial distribution at the same time. 
This method was already successfully applied to astrometric data \citep[i.e.][]{2018A&A...615A..49C, CantatGaudin18gdr2}, where it serendipitously revealed 60 new Milky Way stellar clusters while analysing previously known clusters.

After querying the \gaia\ archive data around the candidates, we apply UPMASK, the workflow of which is outlined below. 

\begin{enumerate}
\item Sample a new dataset drawn from the probability distribution functions defined by the original measurements and their reported covariance matrices. 
\item Create small groups in the parameter space ($\mu_{\alpha*}, \mu_{\delta}, \varpi$) through a \textit{k}-means clustering algorithm \citep{Forgy1965, Lloyd1982}, with a large $k$ with respect to the dataset size, guaranteeing $\sim 10-15$ objects per group. 
\item Test each small group for compatibility with a uniform spatial distribution in $(l,b)$ based on the branch lengths of minimum spanning trees \citep[e.g.][]{GrahamAndHell1985}, after which those that are compatible are discarded as field stars. 
\item Repeat steps (2-3) until no star is discarded, and either the remaining ones are assigned as stellar cluster members at this iteration or all stars are discarded and no cluster is detected. 
\item Repeat steps (1-4) up to the maximum number of iterations (in this study we performed the loop 100 times).
\item Compute a membership score as the frequency with which each star was assigned as a cluster member.
\end{enumerate}

The resulting clustering score is therefore based on the 5D ($\alpha, \delta, \mu_{\alpha*}, \mu_{\delta}, \varpi$) information of each star and associated nominal uncertainties.

\section{Identification of clusters in the Perseus direction} \label{Discussion}

After an automated analysis and human visual inspection of all colour-magnitude diagrams and positional maps of the candidates, we ended up with 133 stellar aggregates that look like potential clusters. 41 are hitherto unreported clusters, and an additional five are known obects that had not yet been identified in the \textit{Gaia}~DR2 data. The location of their members and their colour-magnitude diagrams are displayed in Figs.~\ref{fig:example_COIN-Gaia_1} to \ref{fig:example_Czernik_15}. The full membership list is available in electronic form.

\subsection{Re-identification of known objects}
The majority of the identified aggregates correspond to known clusters. 87 of them were already identified with \gaia~DR2 astrometry by \citet{CantatGaudin18gdr2} in a search that relied on prior information on the location and expected dimension of clusters from the literature, which allowed them to identify 227 objects in the area investigated in the present study. Our unsupervised search did not recover all of those 227, but was however able to recover five objects (Czernik~5, Czernik~15, FSR~0494, FSR~0519, and NGC~886) missed by \citet{CantatGaudin18gdr2}.   
In all five cases, the apparent sizes listed in the catalogues of \citet{Dias2002} and \citet{Kharchenko2013} are too small for the clusters to appear as a clear overdensity in the field of view.

The nearest of these five mischaracterised clusters is NGC~886, an OC in Cassiopeia. 
This cluster was first observed by J. Herschel in 1829 \citep[No.~214 in][]{herschel1833} and is listed under its current name in the original \textit{New General Catalogue of Nebul{\ae} and Clusters of Stars} \citep{Dreyer1888}. Rediscovered as Stock~6 in the 1950s, it was later flagged as ``non existent" in the \textit{Revised New General Catalogue of Nonstellar Astronomical Objects} \citep[RNGC;][]{sulentic1973revised}. The catalogue of \citet{Dias2002} lists Stock~6, with an apparent radius of 0.12$^{\circ}$, but flags it as a dubious grouping, while \citet{Kharchenko2013} lists NGC~886, with a total radius of 0.17$^{\circ}$. 
In this study, we find a clear centrally-concentrated distribution of stars within 0.5$^{\circ}$ of the reported position of NGC~886/Stock~6 (see Fig.~\ref{fig:example_NGC_886}).

\subsection{Newly discovered clusters} \label{stats}
The remaining 41 groups are not (to the best of our knowledge) listed in the literature \citep{Dias2002,Kharchenko2013,Schmeja14,scholz15,Roeser16,Castro18,Ferreira19}. We remark that the colour-magnitude diagrams of most of the newly reported clusters present broadened sequence (see Figs.~\ref{fig:example_COIN-Gaia_1} to \ref{fig:example_COIN-Gaia_41}) and red turn off point ($G_{BP}$-$G_{RP}$$\sim$0.6 to more than 1), wich suggests that they are affected by differential reddening. They are however all clearly visible in astrometric space. Their positions and mean proper motions and parallaxes are given in Table~\ref{table:meanparams}, along with the radius r$_{50}$ containing half the cluster members we identified. We also provide a rough distance from the mean parallax, after correcting for a zero point offset of 0.029\,mas \citep{Lindegren18,Arenou18}.

We perform an additional statistical diagnostic to assess the significance of the signature of the clusters in astrometric space. Working with all stars in a field of view of radius 2$\times$r$_{50}$, we bin stars in square cells of 0.5\,mas\,yr$^{-1}$ in proper motion space and compute the density distribution of parallaxes (using a Gaussian kernel of sigma 0.1\,mas) in each cell\footnote{We only consider cells containing at least 10 stars, in order to mitigate the noise introduced by sparsely populated cells}. We show in the middle right panels of Figs.~\ref{fig:example_COIN-Gaia_1} to \ref{fig:example_Czernik_15} that the parallax distribution in the proper motion bin containing the cluster is always more peaked (denser relative density) than the mean of the other bins. To quantify how the proper motion selection reveals a peaked parallax distribution we calculate the difference between the density in the cluster bin and the mean density of the field bins, divided by the standard deviation of the field bins. This number can be interpreted as a signal-to-noise ratio (S/N) of the signature of the cluster in parallax space.

Reciprocally, we quantify how the density in proper motion space is enhanced when selecting stars in a 0.2\,mas range centred on the cluster parallax, compared to the mean density in bins not containing the cluster. The proper motions of stars (highlighted from their parallax only) are shown in the right panels of Figs.~\ref{fig:example_COIN-Gaia_1} to \ref{fig:example_COIN-Gaia_41}.

\begin{figure}[!htp]
\begin{center}
\includegraphics[width=0.5\textwidth]{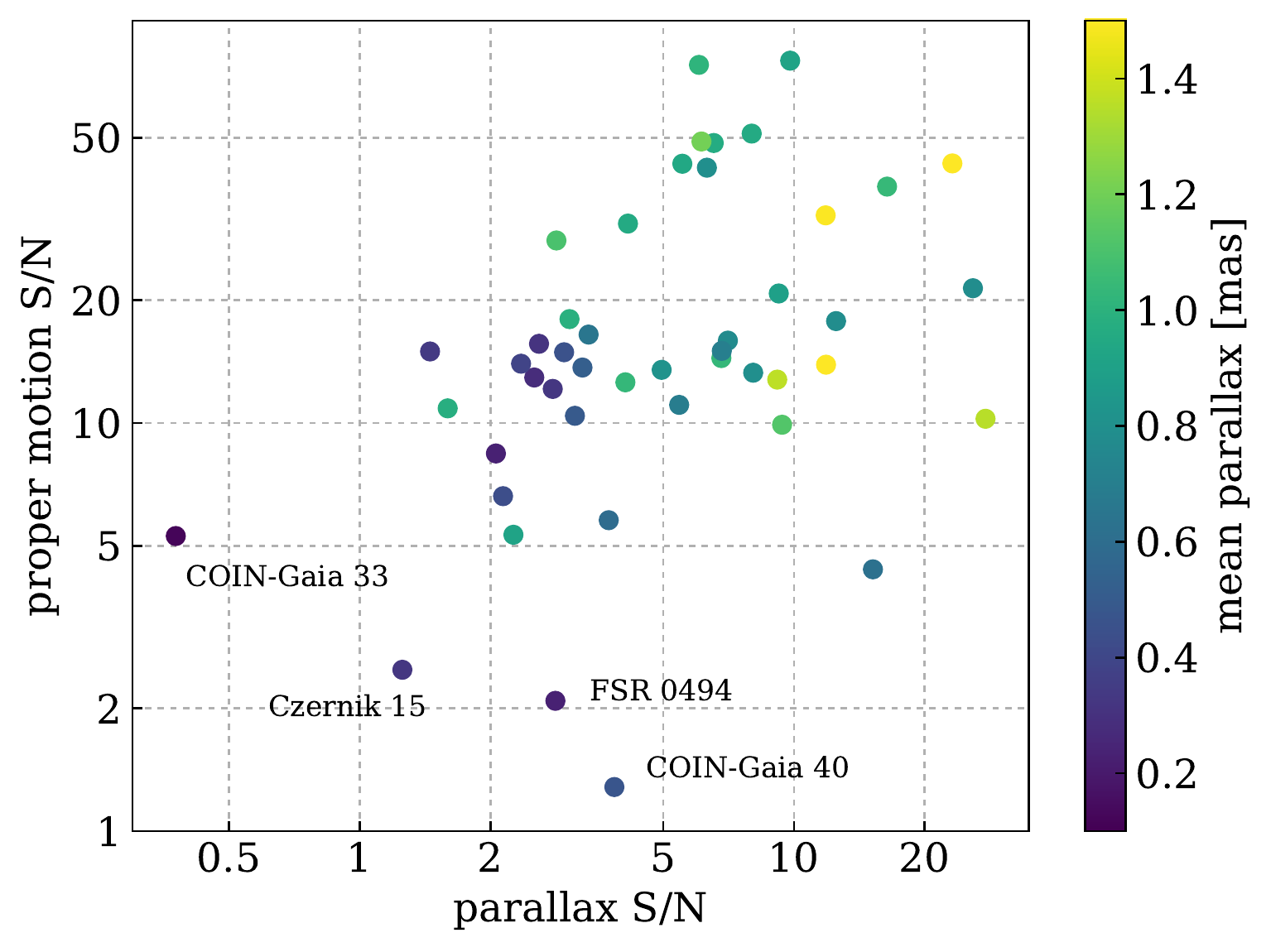}
\caption{Signal in parallax and proper motion space (see Sect.~\ref{stats}) for the 41 newly discovered clusters and five clusters re-identified in this study.
\label{fig:SNratio}}
\end{center}
\end{figure}

In Fig.~\ref{fig:SNratio} we show the S/N ratio in parallax and proper motion space for the 41 clusters discovered in this study and the five clusters we re-identified. Unsurprisingly, the signal in parallax space is stronger for the most nearby clusters. The most distant new cluster (COIN-Gaia~33, $\varpi$$\sim$0.12\,mas) barely stands out as an overdensity in parallax space, but its signature is clearly visible in proper motion space.

Based on their aspect in positional space and the aspect of their colour-magnitude diagram, we divide the 41 new clusters into 28 grade A (most certain candidates) and 13 grade B clusters. Their main parameters (location, apparent size, proper motions and parallax) are listed in Table~\ref{table:meanparams}. The grade B clusters tend to be more distant, more reddened, and their colour-magnitude diagrams are not well-defined, likely due to differential extinction. We remark that the four known objects Czernik~5, Czernik~15, FSR~0494, and FSR~0519 present a similarly blurred colour-magnitude diagram and weak signal in proper motion and parallax space.

\subsection{Consequences for the cluster census}

The unprecedented quality of the astrometry provided by the \gaia\ mission allows for new discoveries of groups of stars sharing a common proper motion and parallax, which is especially powerful when investigating populations projected against a dense background. We have verified that none of the objects characterised in this study are visible as significant overdensities in optical images of the Galactic plane.

The majority of new clusters reported in this work are nearby objects, 33 of which are located within 2~kpc of the Sun.
Their discovery represents a $\sim$50\% increase in the number of clusters identified from \textit{Gaia}~DR2 astrometry in this direction and distance range\footnote{\citet{CantatGaudin18gdr2} have identified 68 clusters in this region within 2\,kpc of the Sun. The present study adds NGC~886 to the list of known clusters in this region and distance range. None of the recent discoveries of \citet{Castro18} and \citet{Ferreira19} fall within this region.}. These 33 new nearby clusters also represent a $\sim$20-25\% increase with respect to the catalogues of \citet{Dias2002} and \citet{Kharchenko2013}. These two widely-used reference catalogues include a significant fraction of objects flagged as putative or dubious, some of which were only very recently found to be asterisms \citep[e.g.][]{Han2016,Kos2018}. Table~\ref{table:numbers} contains a summary of the total number of clusters listed by different authors in the region investigated in the present study.

\begin{table}[h!]
\begin{center}
\caption{ \label{table:numbers} Number of clusters listed by various authors in the region investigated in the present study.}
	\begin{tabular}{ c c c c c }
	\hline
	\hline
distance & DAML$^a$ & MWSC$^b$ & GDR2$^c$ & COIN-Gaia$^d$ \\
	\hline
<1\,kpc	& 30	& 33	& 23		& 14	\\
<2\,kpc	& 140	& 180	& 69		& 33	\\
all	& 363	& 430	& 227		& 41	\\
	\hline
	\hline
	\end{tabular}
	
\tablefoot{ $^a$: from \citet{Dias2002}; $^b$: \citet{Kharchenko2013}; $^c$: known clusters identified from \textit{Gaia}~DR2 astrometry in \citet{CantatGaudin18gdr2} and this study; $^d$: previously unknown clusters first reported in this study.} 
\end{center}

\end{table}

Since our blind search is unable to recover all the known objects within 2\,kpc, it also certainly failed to detect a significant number of unknown OCs that other methods might be able to uncover in the \textit{Gaia}~DR2 data. In particular, the flagging of candidates described in Sect.~\ref{cflagging} is based on proper motions only, and is therefore likely biased towards clusters whose proper motions are significantly different from the field stars.

\begin{figure}[!htp]
\begin{center}
\includegraphics[width=0.45\textwidth]{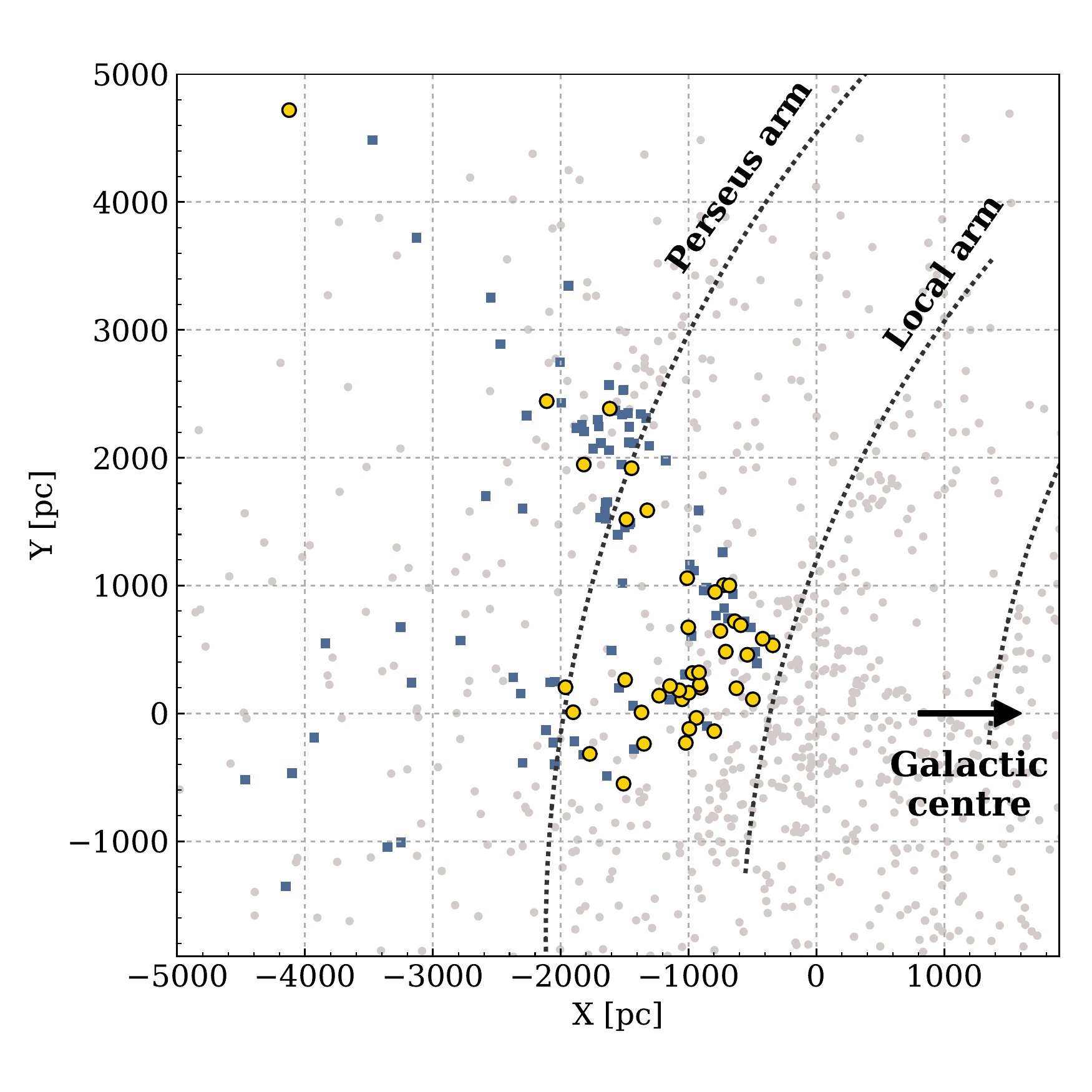}
\caption{Locations of OCs on the projected Galactic plane, for both previously known OCs (gray dots) and new COIN-Gaia open clusters (yellow dots). In addition, blue symbols indicate known OCs that our blind search re-identified. The dotted lines indicate the spiral arm model of \citet{Reid2014}. The Sun is located at coordinates $(0, 0)$.
\label{fig:milkwayprojection}}
\end{center}
\end{figure}

The present study shows that the assumption of completeness often made in OC studies \citep[e.g.][]{Buckner2014, Lin2015, Joshi2016, Matsunaga18, Piskunov2018} needs to be seriously reevaluated in the \textit{Gaia} era. \citet{CantatGaudin18gdr2} have also shown that the samples on which the completeness analyses are based are also highly contaminated by false positives. We note that none of the new objects are located in the gap of the Perseus arm (see Fig.~\ref{fig:milkwayprojection}), and the region $l\in[140;160]$ still appears to be almost devoid of OCs in the distance range $\sim$1-2\,kpc \citep[as already noted by e.g.][]{Vazquez2008}.

Although the \textit{Gaia}-DR2 catalogue represents an unprecedented improvement in the amount and quality of astrometric data available to astronomers, we point out that the next \textit{Gaia} data releases planned for the upcoming years will incrementally refine the parallax and proper motion measurements, allowing us to better discern stellar clusters and possibly to discard some groupings identified in the \textit{Gaia}-DR2 data as false positives.

\begin{figure*}[!htp]
\begin{center}
\includegraphics[width=0.95\textwidth]{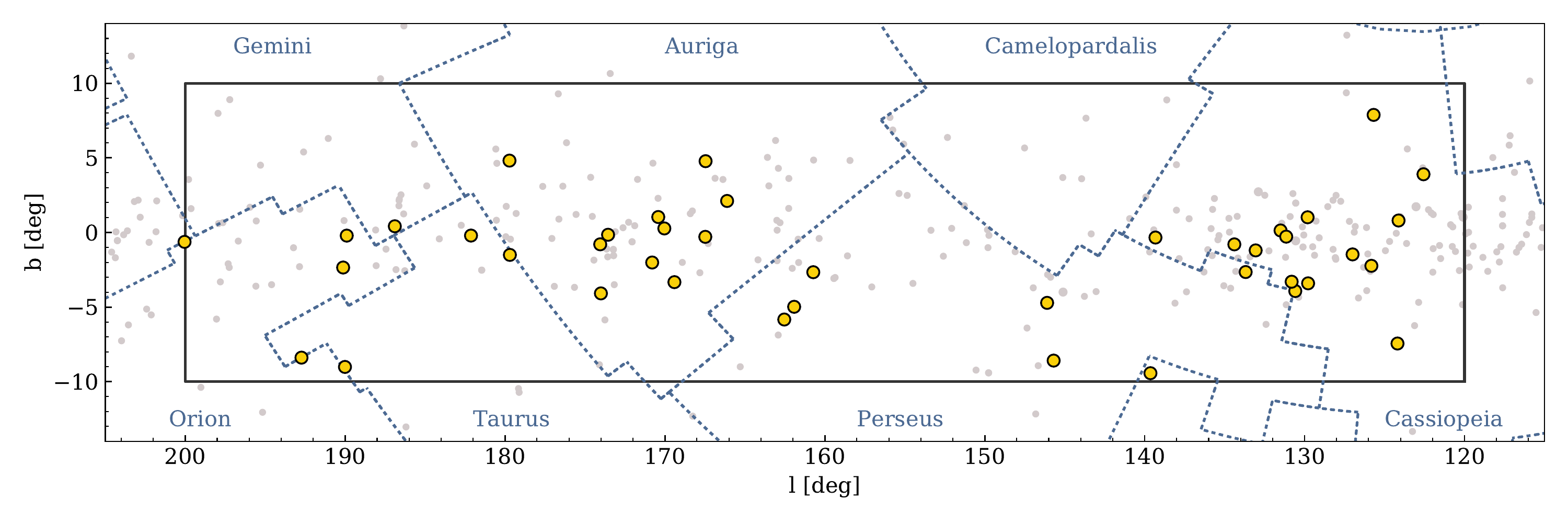}
\caption{Projected Galactic map of previously known OCs (gray points) and the new COIN-Gaia open clusters (yellow dots) detected in this work. 
\label{fig:clustercandidates}}
\end{center}
\end{figure*}

\section{Conclusions}
\label{Conclusions}

This study reports the discovery of 41 new OCs in the direction of Perseus at the Galactic coordinates $l \in [120^\circ; 200^\circ]$ and $b \in [-10^\circ; 10^\circ]$. 
This work employs a fully automated and scalable coarse-to-fine search algorithm, tailored to astrometric catalogues. 
The search is composed of three main steps: a fast density-aware sky tiling, a fast recommendation system for flagging OC candidate regions based on proper motion similarity, and cluster membership assignment and independent validation.

The majority of the 41 new clusters reported in this work are nearby objects, 33 of which are located within 2~kpc of the Sun.
Their discovery represents a significant increase of at least 20\% in the OC population relative to previously known objects in the same region. 
The sample is divided in 28 high certainty OCs and 13 plausible OCs, for which we provide the location, apparent size, and mean astrometric parameters. 
This works challenges the previous claim that the cluster census was complete up to 1.8 kpc, and suggests that many discoveries are still to be made in our nearby galactic environment.

\begin{acknowledgements}
We thank the anonymous referee for the comments and suggestions that helped improve this manuscript.

This work was created during the $\rm 5^{th}$ COIN Residence Program \href{https://cosmostatistics-initiative.org/residence-programs/coin-residence-program-5-chania-greece/}{(CRP\#5)} held in Chania, Greece in September 2018, with support from CNRS and IAASARS.  
We thank Vassilis Charmandaris for encouraging the accomplishment of this event. This project is financially supported by CNRS as part of its MOMENTUM programme over the 2018 – 2020 period.

TCG acknowledges support from Juan de la Cierva - formaci\'on 2015 grant, MINECO (FEDER/UE). 
This work was supported by the MINECO (Spanish Ministry of Economy) through grant ESP2016-80079-C2-1-R (MINECO/FEDER, UE) and MDM-2014-0369 of ICCUB (Unidad de Excelencia 'María de Maeztu').
AKM acknowledges the support from the Portuguese Funda\c c\~ao para a Ci\^encia e a Tecnologia through grants SFRH/BPD/74697/2010, PTDC/FIS-AST/31546/2017, and from the ESA contract AO/1-7836/14/NL/HB. AKM and AM acknowledge the support from the Portuguese Strategic Programme UID/FIS/00099/2013 for CENTRA.
RSS acknowledges the support from  NASA under the Astrophysics Theory Program grant 14-ATP14-0007.
RS has received funding from the European Research Council (ERC) under the European Union's Horizon 2020 research and innovation programme under grant agreement No 771282
AF is supported by a McWilliams Postdoctoral Fellowship.
AIM is a New York University Graduate School of Arts and Sciences Ted Keusseff Fellow advised by David W. Hogg.
BM is a Principal's Career Development Scholar at the University of Edinburgh.
We acknowledge feedback from Madhura Killedar, whose attendance at CRP\#5 was supported by the Sydney Informatics Hub at The University of Sydney.

We acknowledge partial support from the Funda\c c\~ao de Amparo \`a Pesquisa do Estado de S\~ao Paulo grant 2014/13407-4.  This work has made use of the computing facilities of the Laboratory of Astroinformatics (IAG/USP, NAT/Unicsul), whose purchase was made possible by the Brazilian agency FAPESP (grant 2009/54006-4) and the INCT-A, and we thank the entire LAi team, specially Ulisses Manzo Castello and Alex Carciofi for the support.

This work has made use of results from the ESA space mission \gaia, the data from which were processed by the \gaia\ Data Processing and Analysis Consortium (DPAC). 
Funding for the DPAC has been provided by national institutions, in particular the institutions participating in the \gaia\ Multilateral Agreement. 
The \gaia\ mission website is http://www.cosmos.esa.int/gaia. Some of the authors are members of the \gaia\ DPAC. 

The  Cosmostatistics  Initiative\footnote{\href{https://cosmostatistics-initiative.org}{https://cosmostatistics-initiative.org}} (COIN) is a non-profit organization whose aim is to nourish the synergy between astrophysics, cosmology, statistics, and machine learning communities. 
COIN acknowledges the support from the \texttt{Overleaf}\footnote{\href{https://www.overleaf.com}{https://www.overleaf.com}} collaborative platform.

\end{acknowledgements}

\bibliographystyle{aa}
\bibliography{bibliography}

\clearpage

\appendix
\section{COIN-Gaia open clusters}
\label{app_tab}

\begin{table}[h!]
\begin{center}
\onecolumn
\caption{ \label{table:meanparams} Mean parameters of the open clusters characterised in this study.}
	\begin{tabular}{ c  c  c  c  c  c  c | c  c  c  c | c  c  c }
	\hline
	\hline

OC & $\alpha$ & $\delta$ & $l$ & $b$ & r$_{50}$ & $N$ & $\mu_{\alpha}*$ & $\sigma_{\mu_{\alpha}*}$ & $\mu_{\delta}$ & $\sigma_{\mu_{\delta}}$ & $\varpi$ & $\sigma_{\varpi}$ & $d$  \\

  & [deg] & [deg] & [deg] & [deg] & [deg] &   &  \multicolumn{4}{c|}{[mas\,yr$^{-1}$]}  & [mas] & [mas] & [pc] \\
	\hline
	\multicolumn{14}{c}{Grade A clusters} \\
	\hline
COIN-Gaia~1 & 11.933 & 66.769 & 122.566 & 3.9 & 0.357 & 88 & -5.03 & 0.12 & -3.04 & 0.11 & 1.55 & 0.04 & 635  \\
COIN-Gaia~2 & 15.06 & 55.409 & 124.192 & -7.441 & 0.187 & 151 & -4.46 & 0.12 & -1.96 & 0.12 & 0.79 & 0.06 & 1226  \\
COIN-Gaia~3 & 18.739 & 60.505 & 125.825 & -2.235 & 0.126 & 98 & -2.43 & 0.1 & -1.81 & 0.12 & 0.78 & 0.05 & 1243  \\
COIN-Gaia~4 & 26.129 & 58.756 & 129.782 & -3.408 & 0.107 & 66 & -0.95 & 0.07 & -1.04 & 0.08 & 0.45 & 0.03 & 2086  \\
COIN-Gaia~5 & 27.408 & 58.078 & 130.583 & -3.927 & 0.224 & 39 & -2.77 & 0.17 & -0.53 & 0.1 & 1.06 & 0.08 & 916  \\
COIN-Gaia~6 & 28.101 & 58.636 & 130.809 & -3.3 & 0.077 & 132 & -2.35 & 0.11 & -0.44 & 0.14 & 0.28 & 0.06 & 3267  \\
COIN-Gaia~7 & 33.738 & 58.466 & 133.686 & -2.651 & 0.152 & 53 & -1.04 & 0.1 & -1.56 & 0.09 & 0.65 & 0.05 & 1477  \\
COIN-Gaia~8 & 39.048 & 50.013 & 139.64 & -9.43 & 0.312 & 69 & 2.51 & 0.11 & -2.49 & 0.11 & 1.36 & 0.04 & 720  \\
COIN-Gaia~9 & 47.748 & 48.023 & 145.694 & -8.585 & 0.265 & 41 & -1.93 & 0.11 & -2.79 & 0.12 & 1.12 & 0.06 & 871  \\
COIN-Gaia~10 & 68.385 & 40.509 & 161.919 & -4.981 & 0.177 & 41 & 2.02 & 0.09 & -3.37 & 0.11 & 0.94 & 0.04 & 1029  \\
COIN-Gaia~11 & 68.11 & 39.479 & 162.538 & -5.834 & 0.336 & 92 & 3.57 & 0.15 & -5.63 & 0.13 & 1.49 & 0.05 & 661  \\
COIN-Gaia~12 & 79.209 & 41.708 & 166.113 & 2.112 & 0.306 & 90 & 2.62 & 0.13 & -4.66 & 0.09 & 1.03 & 0.05 & 944  \\
COIN-Gaia~13 & 83.186 & 42.087 & 167.459 & 4.776 & 1.003 & 165 & -3.83 & 0.18 & -1.66 & 0.17 & 1.93 & 0.09 & 511  \\
COIN-Gaia~14 & 77.696 & 39.195 & 167.476 & -0.293 & 0.157 & 24 & 1.33 & 0.09 & -6.57 & 0.08 & 1.04 & 0.03 & 933  \\
COIN-Gaia~15 & 76.09 & 35.831 & 169.408 & -3.328 & 0.147 & 134 & 0.44 & 0.15 & -3.75 & 0.1 & 0.82 & 0.07 & 1172  \\
COIN-Gaia~16 & 80.179 & 37.438 & 170.038 & 0.27 & 0.059 & 49 & 1.26 & 0.14 & -3.74 & 0.14 & 0.62 & 0.06 & 1535  \\
COIN-Gaia~17 & 81.244 & 37.558 & 170.418 & 1.035 & 0.151 & 83 & 0.36 & 0.16 & -4.28 & 0.12 & 0.89 & 0.07 & 1094  \\
COIN-Gaia~18 & 78.408 & 35.498 & 170.797 & -2.014 & 0.239 & 59 & 0.77 & 0.12 & -4.92 & 0.14 & 0.95 & 0.07 & 1018  \\
COIN-Gaia~19 & 82.188 & 34.29 & 173.556 & -0.153 & 0.172 & 88 & -1.48 & 0.13 & -4.64 & 0.09 & 0.77 & 0.06 & 1245  \\
COIN-Gaia~20 & 78.634 & 31.691 & 174.004 & -4.082 & 0.258 & 55 & 0.55 & 0.14 & -1.45 & 0.09 & 0.91 & 0.05 & 1063  \\
COIN-Gaia~21 & 84.766 & 28.402 & 179.696 & -1.504 & 0.098 & 42 & -0.12 & 0.09 & -3.82 & 0.08 & 0.7 & 0.04 & 1378  \\
COIN-Gaia~22 & 91.06 & 31.602 & 179.721 & 4.815 & 0.086 & 104 & -0.71 & 0.15 & -3.28 & 0.11 & 0.49 & 0.07 & 1927  \\
COIN-Gaia~23 & 87.449 & 27.008 & 182.127 & -0.209 & 0.275 & 108 & -0.32 & 0.16 & -0.96 & 0.13 & 1.03 & 0.08 & 942  \\
COIN-Gaia~24 & 90.693 & 23.203 & 186.893 & 0.416 & 0.194 & 70 & 2.54 & 0.1 & -2.97 & 0.09 & 0.96 & 0.05 & 1006  \\
COIN-Gaia~25 & 91.691 & 20.276 & 189.899 & -0.211 & 0.291 & 112 & -0.5 & 0.15 & -2.65 & 0.12 & 1.2 & 0.07 & 813  \\
COIN-Gaia~26 & 83.771 & 15.721 & 190.008 & -9.01 & 0.107 & 81 & 0.27 & 0.11 & -2.39 & 0.11 & 0.69 & 0.06 & 1395  \\
COIN-Gaia~27 & 85.76 & 13.743 & 192.728 & -8.387 & 0.195 & 78 & 0.7 & 0.13 & -3.6 & 0.12 & 0.91 & 0.05 & 1063  \\
COIN-Gaia~28 & 96.333 & 11.159 & 200.041 & -0.63 & 0.115 & 124 & -1.12 & 0.15 & -0.94 & 0.15 & 0.59 & 0.07 & 1616  \\
	\hline
	\multicolumn{14}{c}{Grade B clusters} \\
	\hline
COIN-Gaia~29 & 15.548 & 63.648 & 124.125 & 0.801 & 0.177 & 29 & -2.65 & 0.05 & -0.32 & 0.04 & 0.31 & 0.03 & 2923  \\
COIN-Gaia~30 & 21.08 & 70.574 & 125.684 & 7.878 & 0.254 & 92 & -6.14 & 0.11 & 2.11 & 0.16 & 1.35 & 0.05 & 727  \\
COIN-Gaia~31 & 21.307 & 61.135 & 127.0 & -1.469 & 0.093 & 37 & -1.98 & 0.07 & -0.53 & 0.05 & 0.38 & 0.04 & 2436  \\
COIN-Gaia~32 & 28.194 & 63.066 & 129.813 & 1.016 & 0.139 & 24 & 0.05 & 0.07 & -2.2 & 0.05 & 0.78 & 0.03 & 1245  \\
COIN-Gaia~33 & 30.276 & 61.475 & 131.15 & -0.28 & 0.05 & 25 & -1.31 & 0.03 & -0.08 & 0.07 & 0.12 & 0.04 & 6576  \\
COIN-Gaia~34 & 31.231 & 61.776 & 131.504 & 0.133 & 0.183 & 47 & -2.23 & 0.16 & 0.88 & 0.09 & 1.01 & 0.06 & 966  \\
COIN-Gaia~35 & 33.472 & 60.039 & 133.06 & -1.201 & 0.043 & 31 & -0.53 & 0.07 & -0.39 & 0.08 & 0.34 & 0.04 & 2702  \\
COIN-Gaia~36 & 36.256 & 59.975 & 134.394 & -0.798 & 0.105 & 7 & -0.97 & 0.06 & -0.56 & 0.04 & 0.43 & 0.01 & 2173  \\
COIN-Gaia~37 & 45.377 & 58.329 & 139.326 & -0.341 & 0.845 & 29 & 0.85 & 0.09 & -2.04 & 0.05 & 0.98 & 0.03 & 995  \\
COIN-Gaia~38 & 51.472 & 51.072 & 146.101 & -4.718 & 0.132 & 73 & 2.03 & 0.14 & -6.82 & 0.11 & 0.79 & 0.05 & 1217  \\
COIN-Gaia~39 & 69.612 & 42.95 & 160.724 & -2.663 & 0.244 & 32 & 0.24 & 0.11 & -2.46 & 0.08 & 0.99 & 0.04 & 980  \\
COIN-Gaia~40 & 81.874 & 33.526 & 174.048 & -0.794 & 0.073 & 28 & 0.39 & 0.06 & -2.73 & 0.05 & 0.47 & 0.04 & 2003  \\
COIN-Gaia~41 & 89.832 & 19.028 & 190.123 & -2.349 & 0.15 & 85 & -0.32 & 0.1 & -3.66 & 0.1 & 0.52 & 0.07 & 1820  \\
	\hline
	\multicolumn{14}{c}{Known clusters} \\
	\hline
FSR~0494 & 6.416 & 63.754 & 120.087 & 1.025 & 0.056 & 72 & -2.5 & 0.11 & -0.85 & 0.07 & 0.23 & 0.05 & 3943  \\
FSR~0519 & 13.092 & 64.596 & 123.032 & 1.724 & 0.085 & 16 & -2.42 & 0.03 & -0.33 & 0.05 & 0.32 & 0.01 & 2833  \\
Czernik~5 & 28.927 & 61.355 & 130.557 & -0.562 & 0.036 & 74 & -2.02 & 0.08 & 0.45 & 0.11 & 0.23 & 0.07 & 3882  \\
NGC~886 & 35.898 & 63.8 & 132.893 & 2.725 & 0.34 & 174 & 3.19 & 0.1 & -4.0 & 0.11 & 0.95 & 0.04 & 1017  \\
Czernik~15 & 50.781 & 52.223 & 145.105 & -3.995 & 0.111 & 36 & 0.33 & 0.09 & -1.07 & 0.09 & 0.32 & 0.04 & 2853  \\

	\hline
	\hline
	\end{tabular}
	
\tablefoot{r$_{50}$: radius containing half the members identified in this study. $N$: Number of stars with frequentist  membership probability $\geq$ 50\%. 
$d$: mode of the distance likelihood after adding +0.029\,mas to the measured parallaxes.  
} 
\twocolumn	
\end{center}

\end{table}

\clearpage

\section{Grade A clusters}
\label{app_A}

\begin{figure*}[ht]
\begin{center} \resizebox{\hsize}{!}{\includegraphics[scale=0.5]{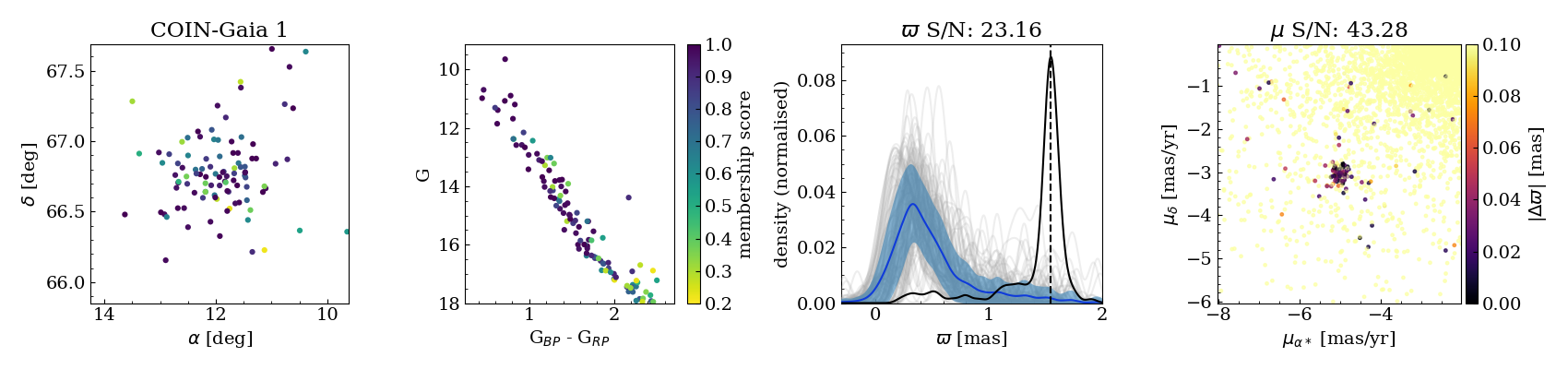}} \caption{\label{fig:example_COIN-Gaia_1} Left: spatial distribution of the stars of COIN-Gaia~1 with membership probabilities $p>$20\%. Middle left: colour-magnitude diagram of the same stars. Middle right: parallax distribution density for stars binned by proper motion (see Sect.~\ref{stats}). Right: Proper motion of all stars in the field of view, colour-coded by their difference in parallax with the value listed in Table~\ref{app_tab}.} \end{center}
\end{figure*}

\begin{figure*}[ht]
\begin{center} \resizebox{\hsize}{!}{\includegraphics[scale=0.5]{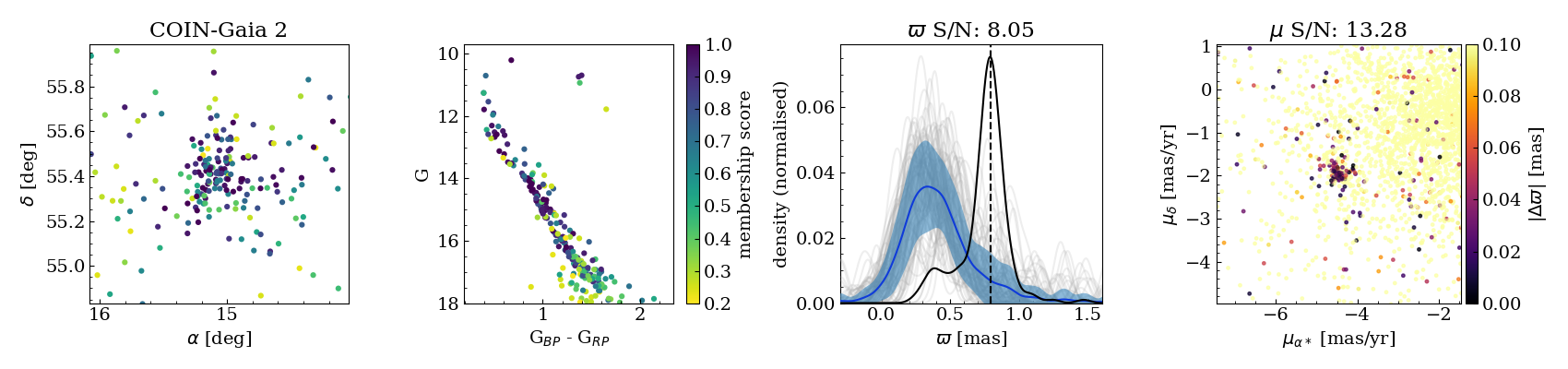}} \caption{\label{fig:example_COIN-Gaia_2} Same as Fig.~\ref{fig:example_COIN-Gaia_1} for COIN-Gaia~2.} \end{center}
\end{figure*}

\begin{figure*}[ht]
\begin{center} \resizebox{\hsize}{!}{\includegraphics[scale=0.5]{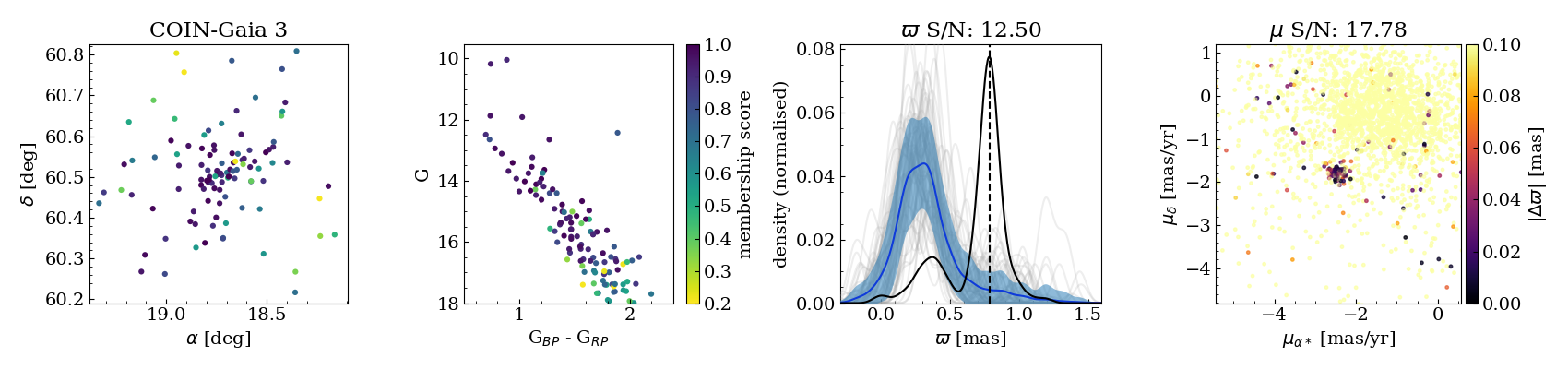}} \caption{\label{fig:example_COIN-Gaia_3} Same as Fig.~\ref{fig:example_COIN-Gaia_1} for COIN-Gaia~3.} \end{center}
\end{figure*}

\begin{figure*}[ht]
\begin{center} \resizebox{\hsize}{!}{\includegraphics[scale=0.5]{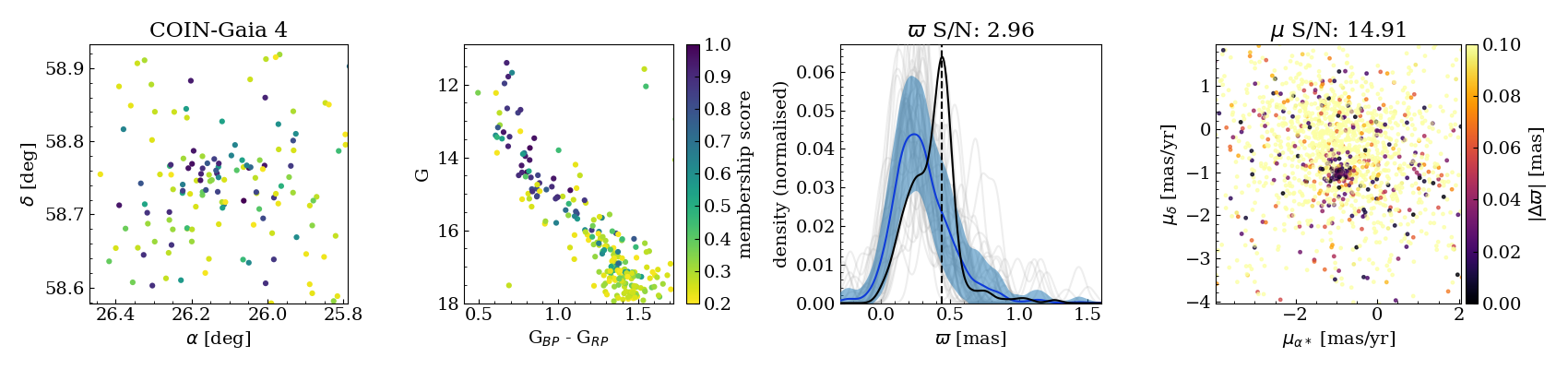}} \caption{\label{fig:example_COIN-Gaia_4} Same as Fig.~\ref{fig:example_COIN-Gaia_1} for COIN-Gaia~4.} \end{center}
\end{figure*}

\begin{figure*}[ht]
\begin{center} \resizebox{\hsize}{!}{\includegraphics[scale=0.5]{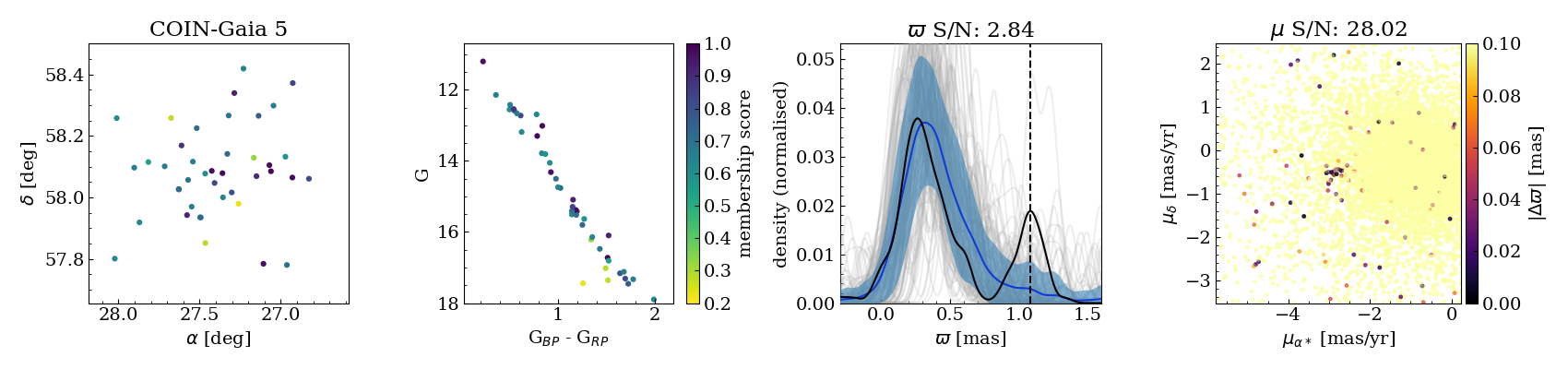}} \caption{\label{fig:example_COIN-Gaia_5} Same as Fig.~\ref{fig:example_COIN-Gaia_1} for COIN-Gaia~5.} \end{center}
\end{figure*}

\begin{figure*}[ht]
\begin{center} \resizebox{\hsize}{!}{\includegraphics[scale=0.5]{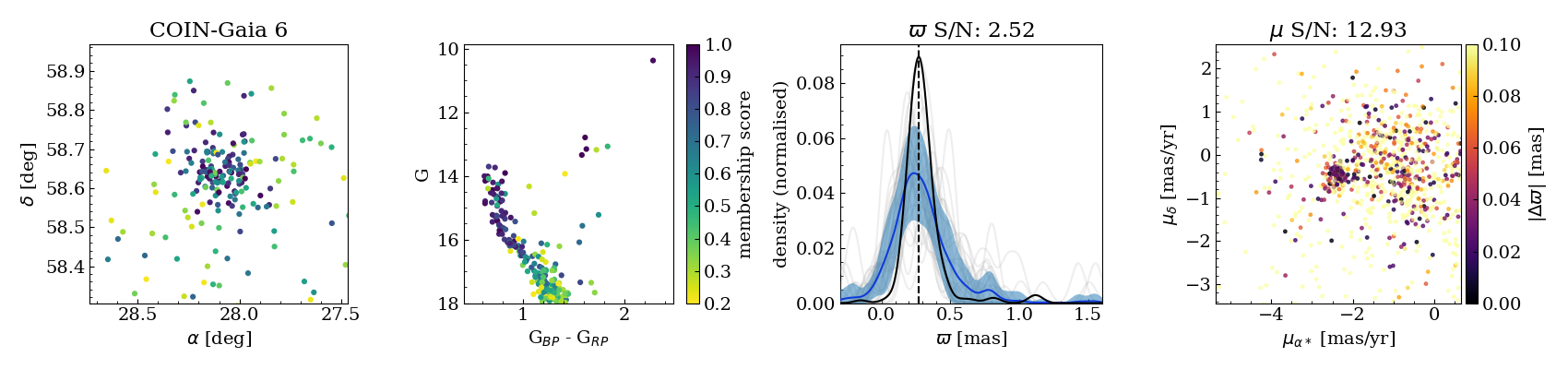}} \caption{\label{fig:example_COIN-Gaia_6} Same as Fig.~\ref{fig:example_COIN-Gaia_1} for COIN-Gaia~6.} \end{center}
\end{figure*}

\begin{figure*}[ht]
\begin{center} \resizebox{\hsize}{!}{\includegraphics[scale=0.5]{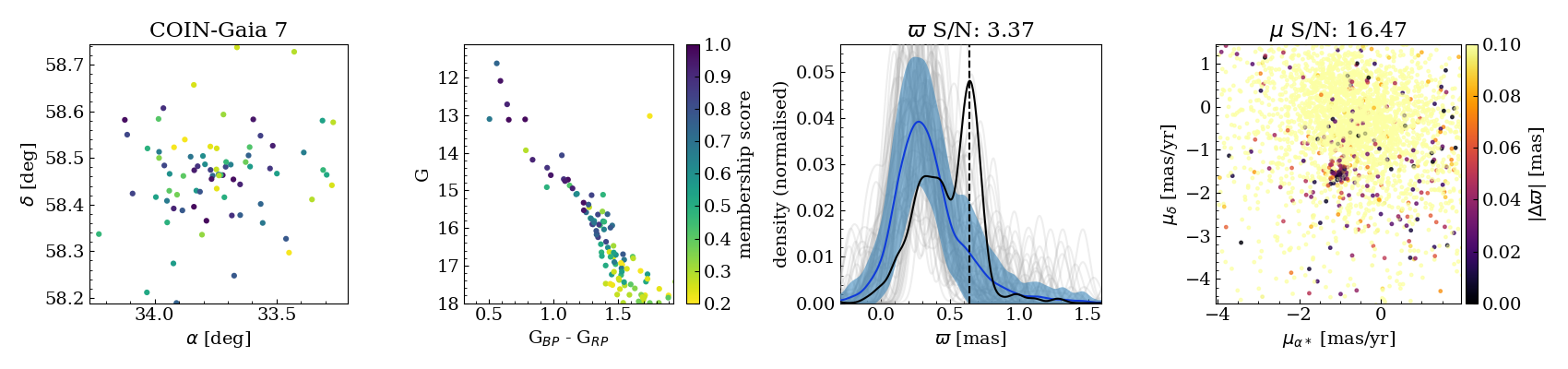}} \caption{\label{fig:example_COIN-Gaia_7} Same as Fig.~\ref{fig:example_COIN-Gaia_1} for COIN-Gaia~7.} \end{center}
\end{figure*}

\begin{figure*}[ht]
\begin{center} \resizebox{\hsize}{!}{\includegraphics[scale=0.5]{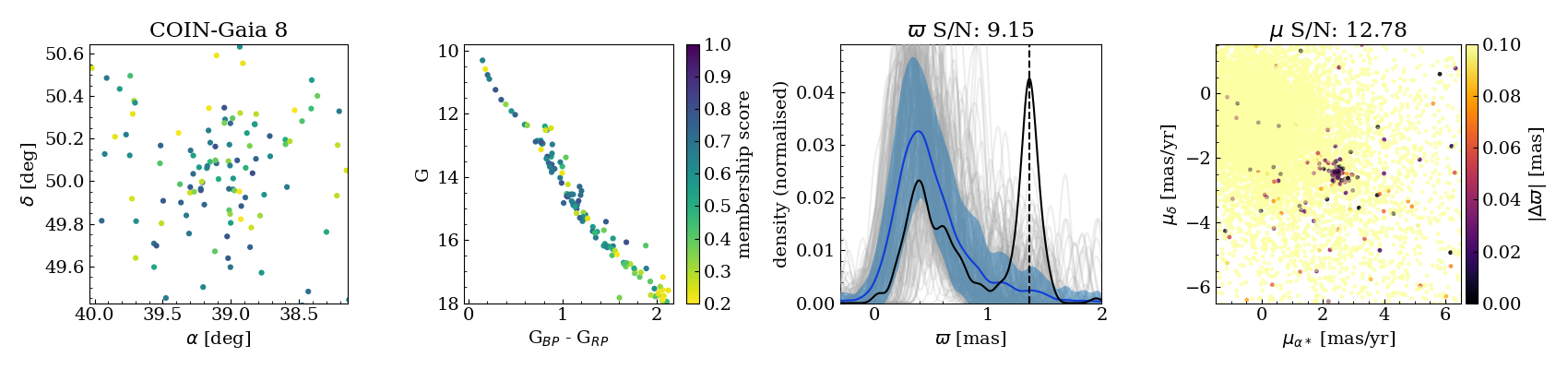}} \caption{\label{fig:example_COIN-Gaia_8} Same as Fig.~\ref{fig:example_COIN-Gaia_1} for COIN-Gaia~8.} \end{center}
\end{figure*}

\begin{figure*}[ht]
\begin{center} \resizebox{\hsize}{!}{\includegraphics[scale=0.5]{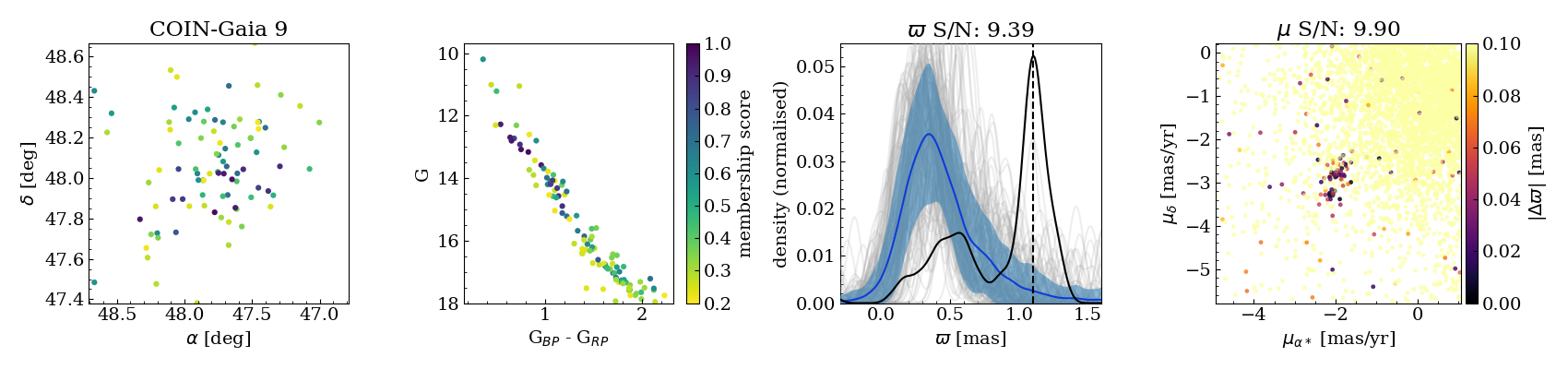}} \caption{\label{fig:example_COIN-Gaia_9} Same as Fig.~\ref{fig:example_COIN-Gaia_1} for COIN-Gaia~9.} \end{center}
\end{figure*}

\begin{figure*}[ht]
\begin{center} \resizebox{\hsize}{!}{\includegraphics[scale=0.5]{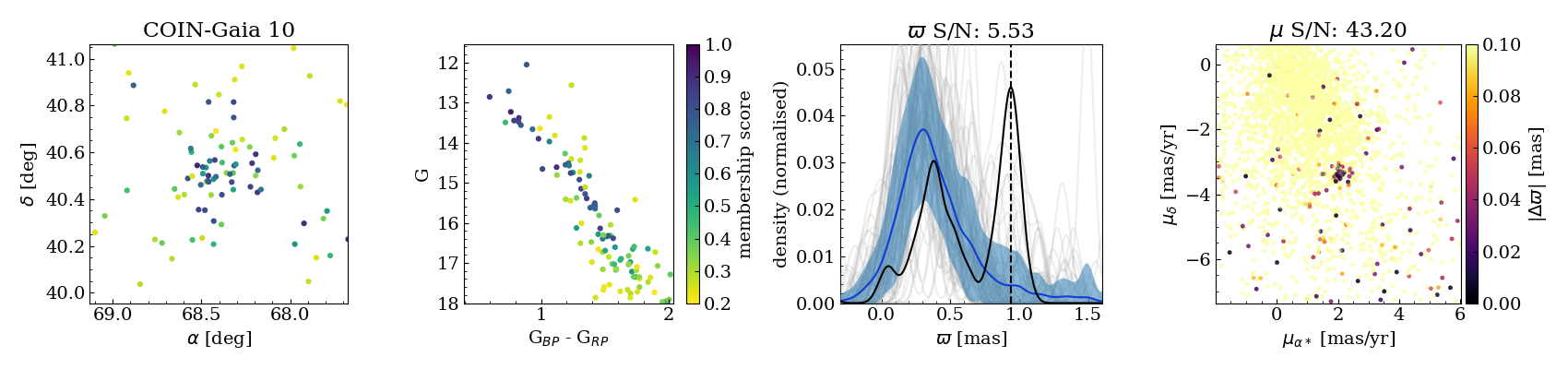}} \caption{\label{fig:example_COIN-Gaia_10} Same as Fig.~\ref{fig:example_COIN-Gaia_1} for COIN-Gaia~10.} \end{center}
\end{figure*}

\begin{figure*}[ht]
\begin{center} \resizebox{\hsize}{!}{\includegraphics[scale=0.5]{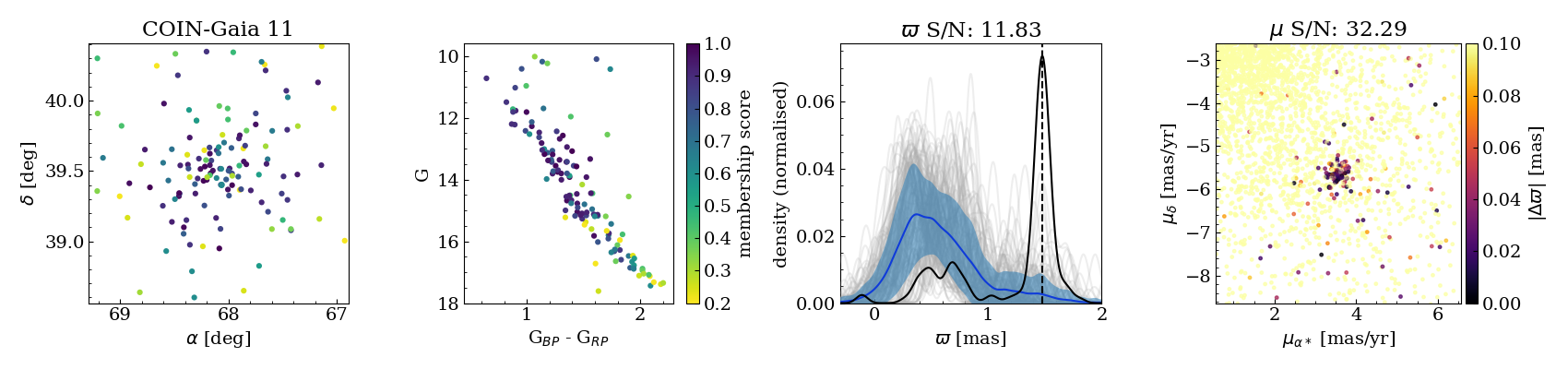}} \caption{\label{fig:example_COIN-Gaia_11} Same as Fig.~\ref{fig:example_COIN-Gaia_1} for COIN-Gaia~11.} \end{center}
\end{figure*}

\begin{figure*}[ht]
\begin{center} \resizebox{\hsize}{!}{\includegraphics[scale=0.5]{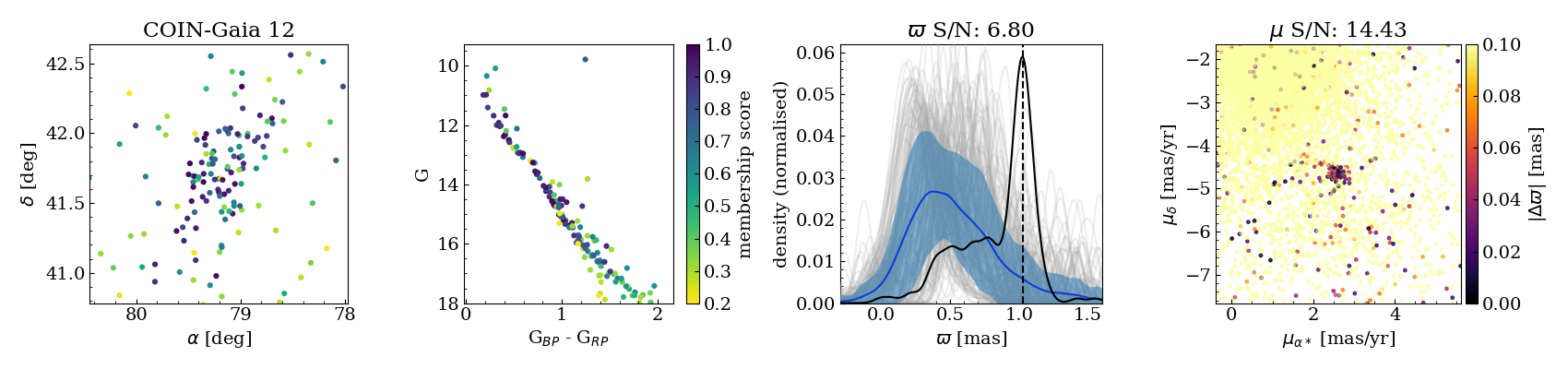}} \caption{\label{fig:example_COIN-Gaia_12} Same as Fig.~\ref{fig:example_COIN-Gaia_1} for COIN-Gaia~12.} \end{center}
\end{figure*}

\clearpage

\begin{figure*}[ht]
\begin{center} \resizebox{\hsize}{!}{\includegraphics[scale=0.5]{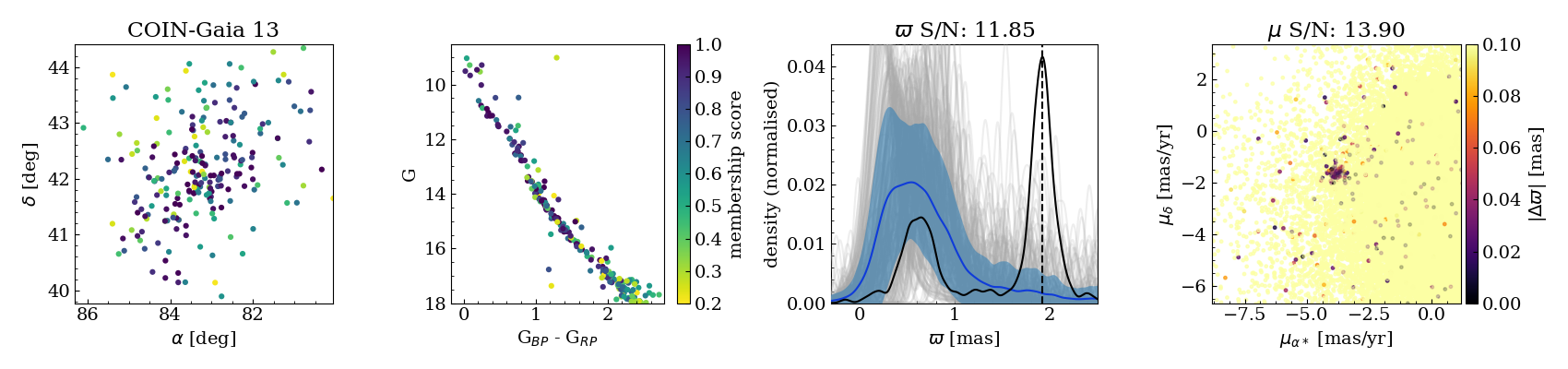}} \caption{\label{fig:example_COIN-Gaia_13} Same as Fig.~\ref{fig:example_COIN-Gaia_1} for COIN-Gaia~13.} \end{center}
\end{figure*}

\begin{figure*}[ht]
\begin{center} \resizebox{\hsize}{!}{\includegraphics[scale=0.5]{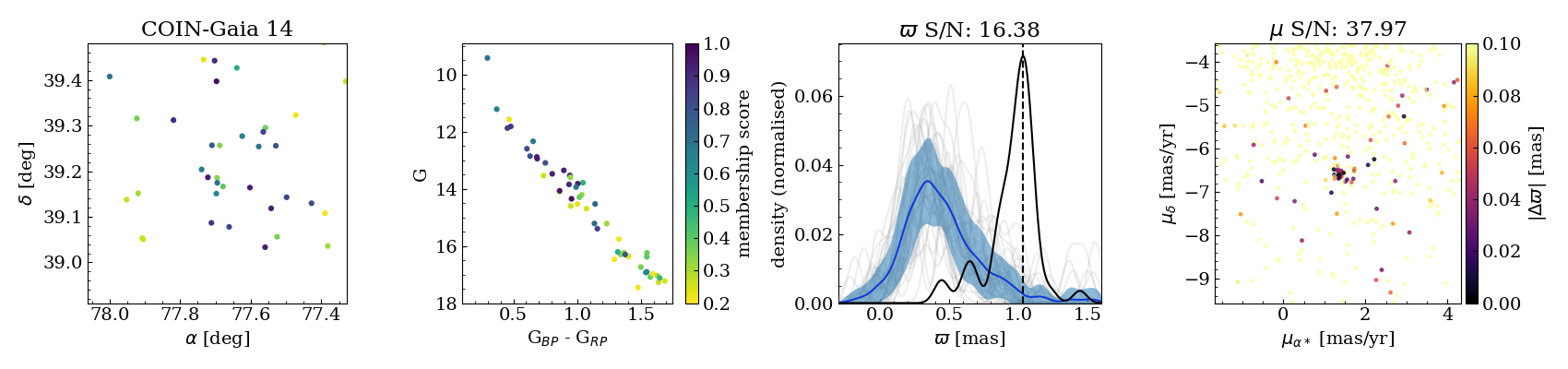}} \caption{\label{fig:example_COIN-Gaia_14} Same as Fig.~\ref{fig:example_COIN-Gaia_1} for COIN-Gaia~14.} \end{center}
\end{figure*}

\begin{figure*}[ht]
\begin{center} \resizebox{\hsize}{!}{\includegraphics[scale=0.5]{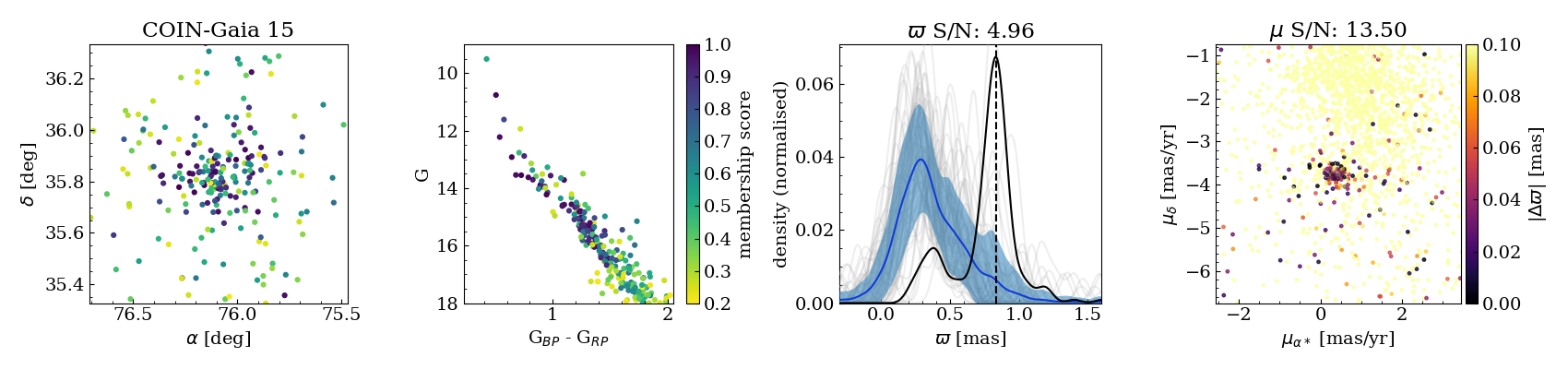}} \caption{\label{fig:example_COIN-Gaia_15} Same as Fig.~\ref{fig:example_COIN-Gaia_1} for COIN-Gaia~15.} \end{center}
\end{figure*}

\begin{figure*}[ht]
\begin{center} \resizebox{\hsize}{!}{\includegraphics[scale=0.5]{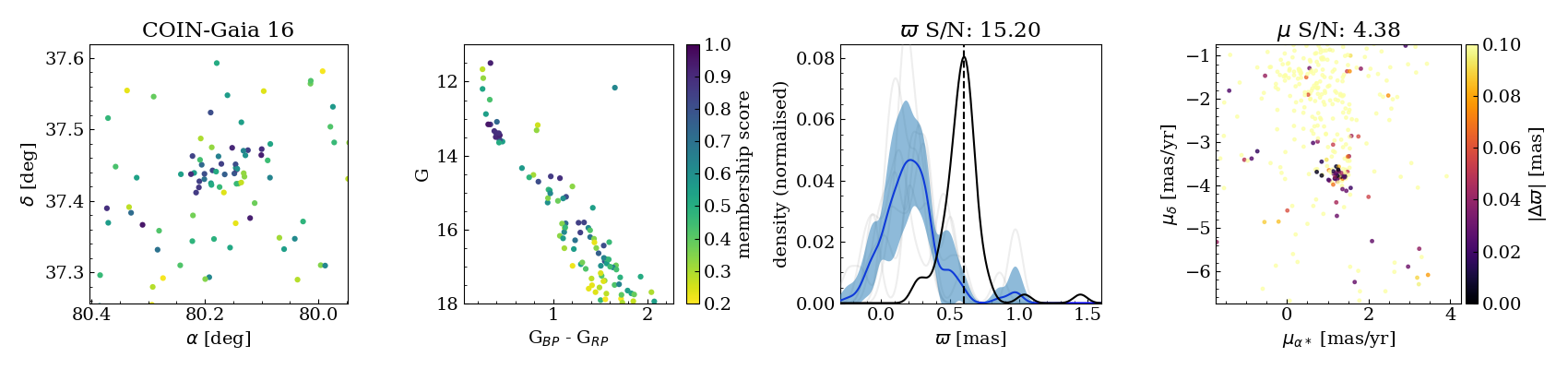}} \caption{\label{fig:example_COIN-Gaia_16} Same as Fig.~\ref{fig:example_COIN-Gaia_1} for COIN-Gaia~16.} \end{center}
\end{figure*}

\begin{figure*}[ht]
\begin{center} \resizebox{\hsize}{!}{\includegraphics[scale=0.5]{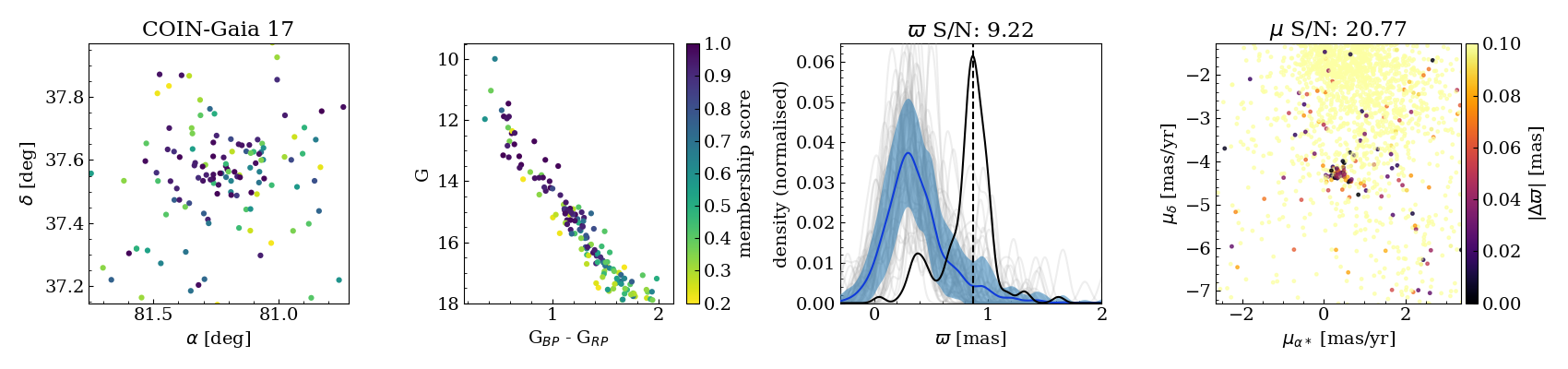}} \caption{\label{fig:example_COIN-Gaia_17} Same as Fig.~\ref{fig:example_COIN-Gaia_1} for COIN-Gaia~17.} \end{center}
\end{figure*}

\begin{figure*}[ht]
\begin{center} \resizebox{\hsize}{!}{\includegraphics[scale=0.5]{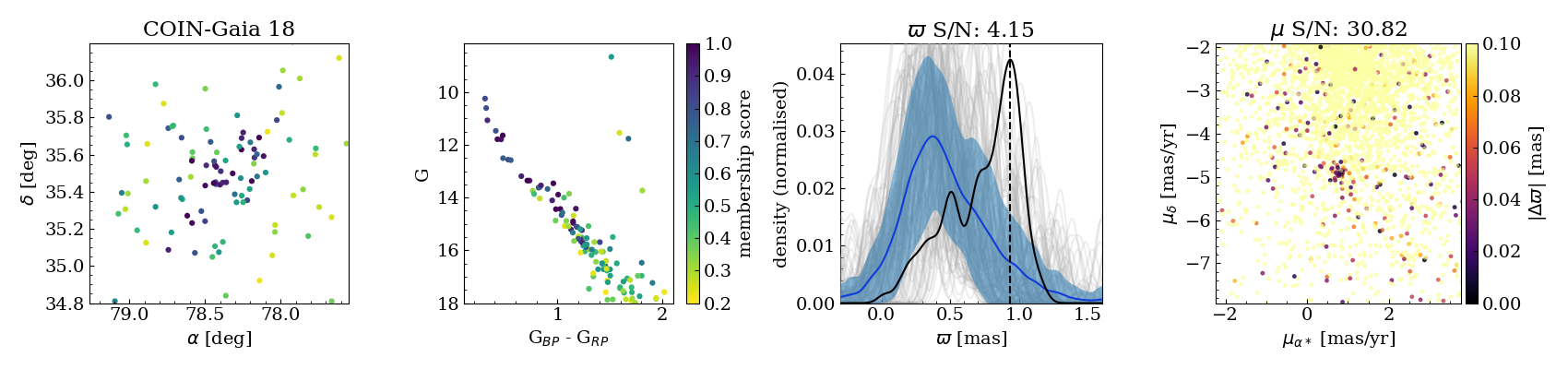}} \caption{\label{fig:example_COIN-Gaia_18} Same as Fig.~\ref{fig:example_COIN-Gaia_1} for COIN-Gaia~18.} \end{center}
\end{figure*}

\begin{figure*}[ht]
\begin{center} \resizebox{\hsize}{!}{\includegraphics[scale=0.5]{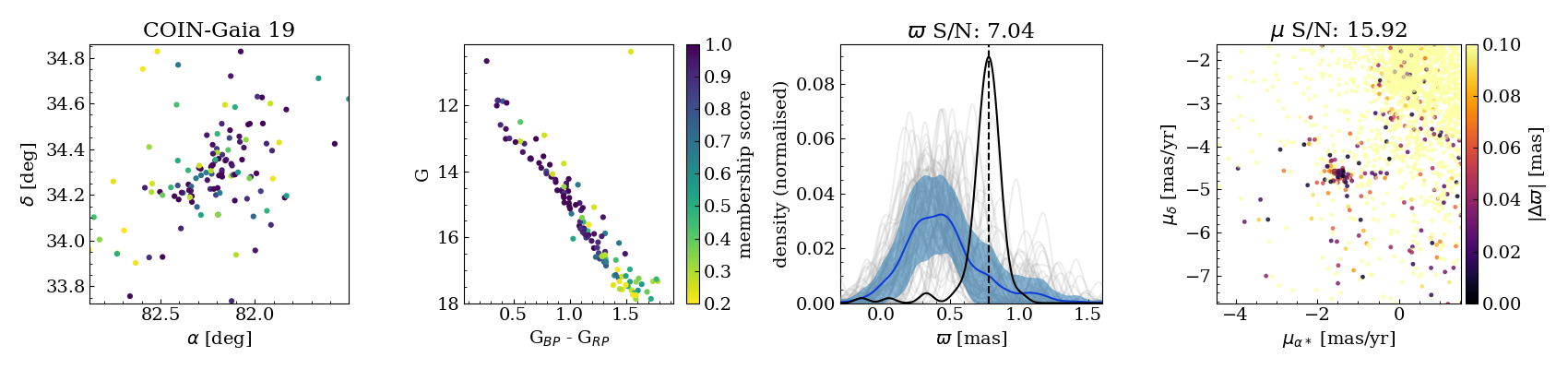}} \caption{\label{fig:example_COIN-Gaia_19} Same as Fig.~\ref{fig:example_COIN-Gaia_1} for COIN-Gaia~19.} \end{center}
\end{figure*}

\begin{figure*}[ht]
\begin{center} \resizebox{\hsize}{!}{\includegraphics[scale=0.5]{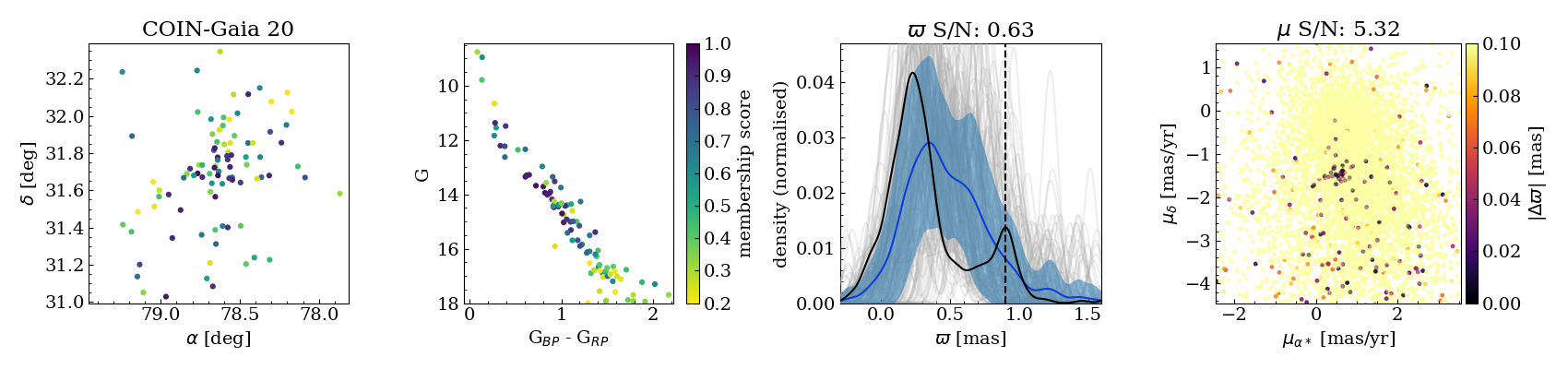}} \caption{\label{fig:example_COIN-Gaia_20} Same as Fig.~\ref{fig:example_COIN-Gaia_1} for COIN-Gaia~20.} \end{center}
\end{figure*}

\begin{figure*}[ht]
\begin{center} \resizebox{\hsize}{!}{\includegraphics[scale=0.5]{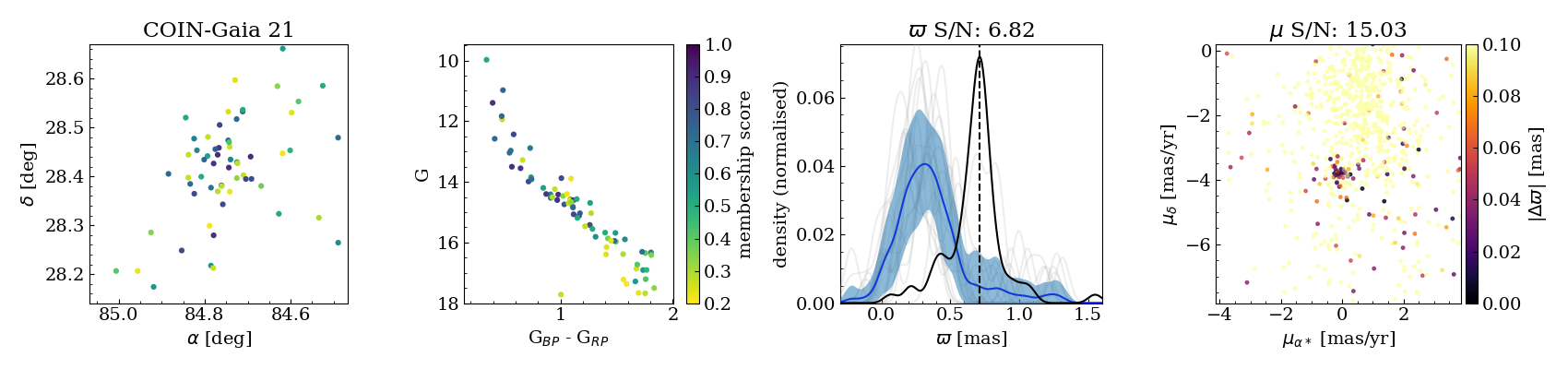}} \caption{\label{fig:example_COIN-Gaia_21} Same as Fig.~\ref{fig:example_COIN-Gaia_1} for COIN-Gaia~21.} \end{center}
\end{figure*}

\begin{figure*}[ht]
\begin{center} \resizebox{\hsize}{!}{\includegraphics[scale=0.5]{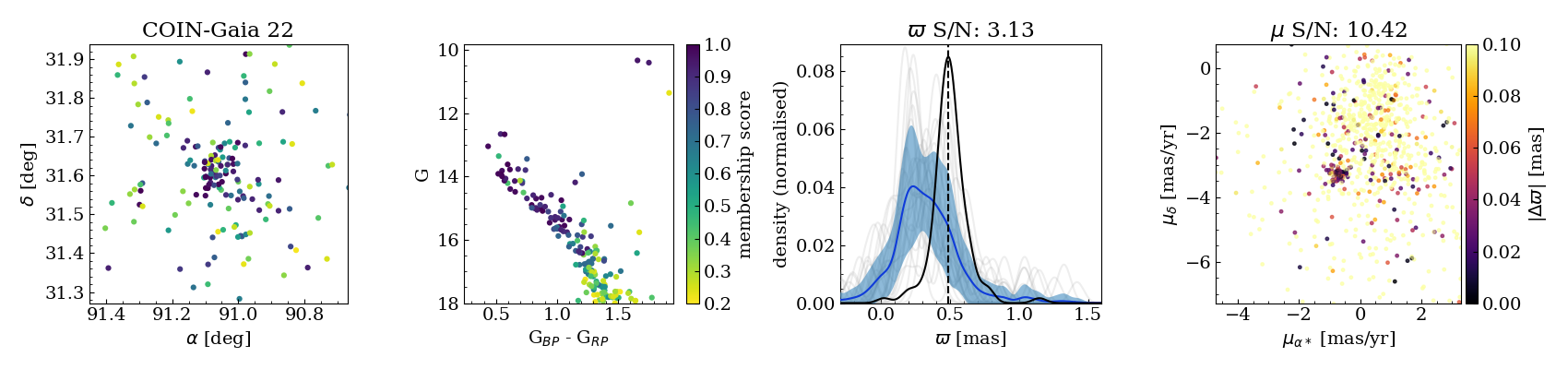}} \caption{\label{fig:example_COIN-Gaia_22} Same as Fig.~\ref{fig:example_COIN-Gaia_1} for COIN-Gaia~22.} \end{center}
\end{figure*}

\begin{figure*}[ht]
\begin{center} \resizebox{\hsize}{!}{\includegraphics[scale=0.5]{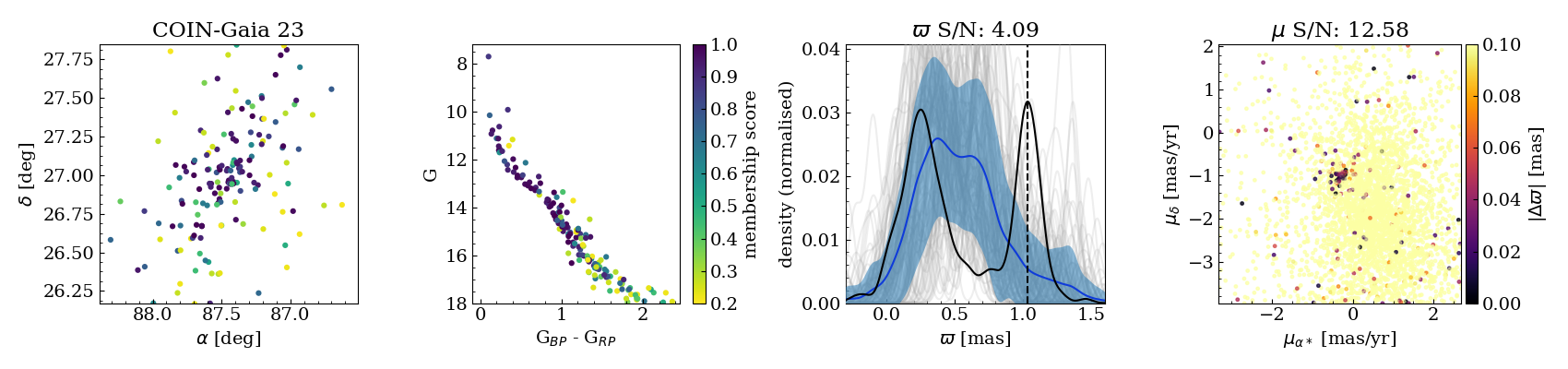}} \caption{\label{fig:example_COIN-Gaia_23} Same as Fig.~\ref{fig:example_COIN-Gaia_1} for COIN-Gaia~23.} \end{center}
\end{figure*}

\begin{figure*}[ht]
\begin{center} \resizebox{\hsize}{!}{\includegraphics[scale=0.5]{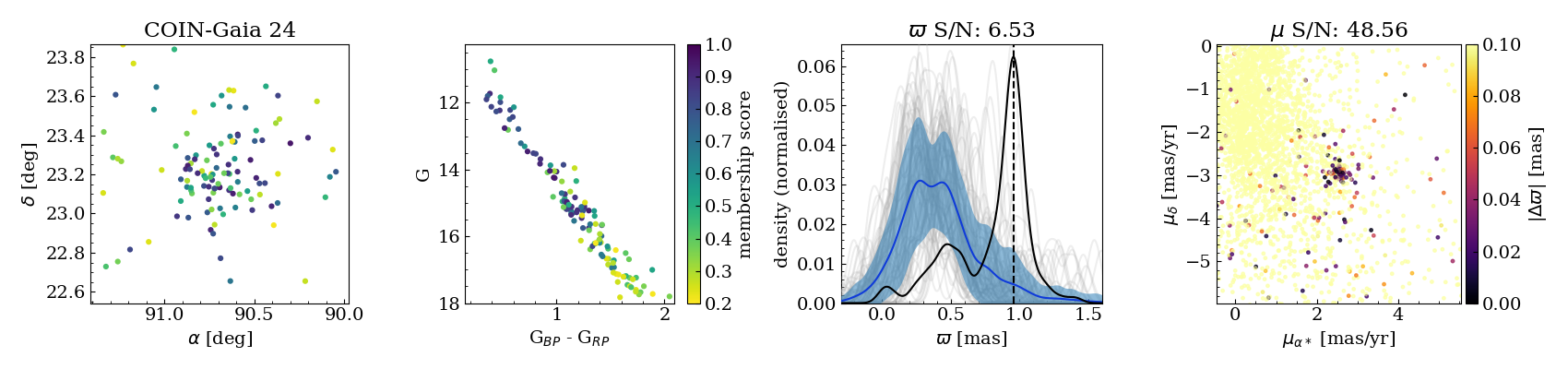}} \caption{\label{fig:example_COIN-Gaia_24} Same as Fig.~\ref{fig:example_COIN-Gaia_1} for COIN-Gaia~24.} \end{center}
\end{figure*}

\begin{figure*}[ht]
\begin{center} \resizebox{\hsize}{!}{\includegraphics[scale=0.5]{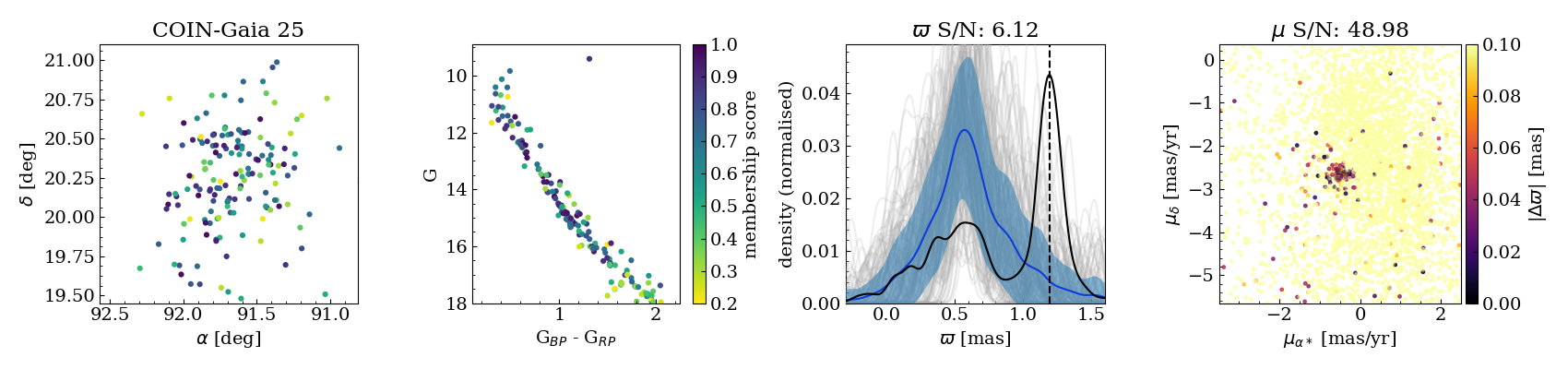}} \caption{\label{fig:example_COIN-Gaia_25} Same as Fig.~\ref{fig:example_COIN-Gaia_1} for COIN-Gaia~25.} \end{center}
\end{figure*}

\begin{figure*}[ht]
\begin{center} \resizebox{\hsize}{!}{\includegraphics[scale=0.5]{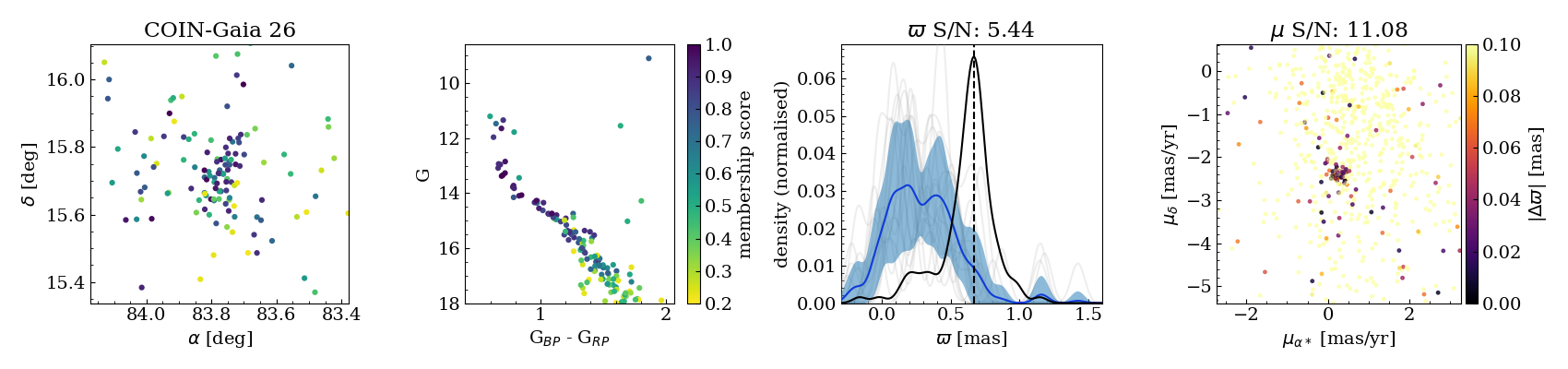}} \caption{\label{fig:example_COIN-Gaia_26} Same as Fig.~\ref{fig:example_COIN-Gaia_1} for COIN-Gaia~26.} \end{center}
\end{figure*}

\begin{figure*}[ht]
\begin{center} \resizebox{\hsize}{!}{\includegraphics[scale=0.5]{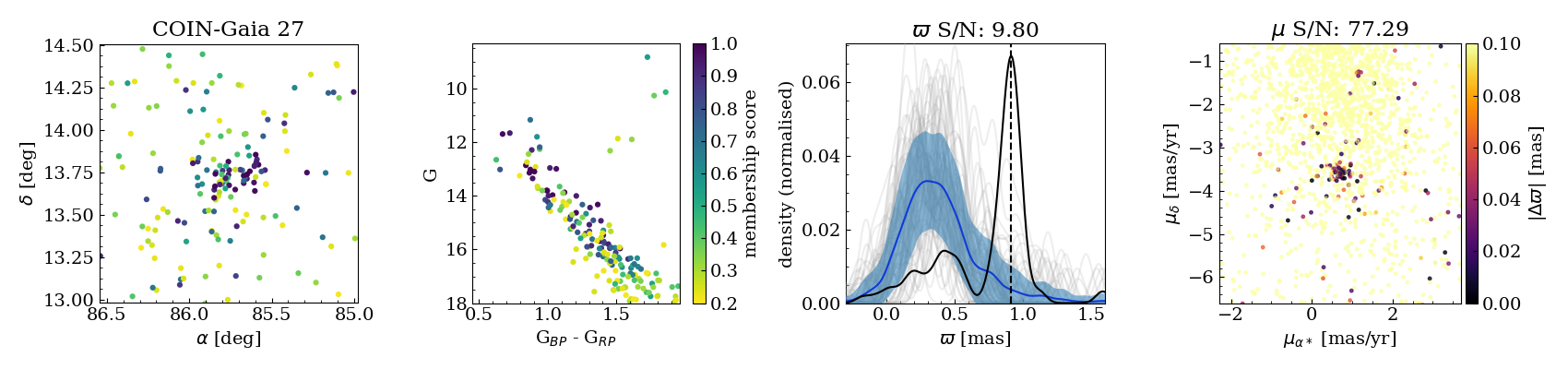}} \caption{\label{fig:example_COIN-Gaia_27} Same as Fig.~\ref{fig:example_COIN-Gaia_1} for COIN-Gaia~27.} \end{center}
\end{figure*}

\begin{figure*}[ht]
\begin{center} \resizebox{\hsize}{!}{\includegraphics[scale=0.5]{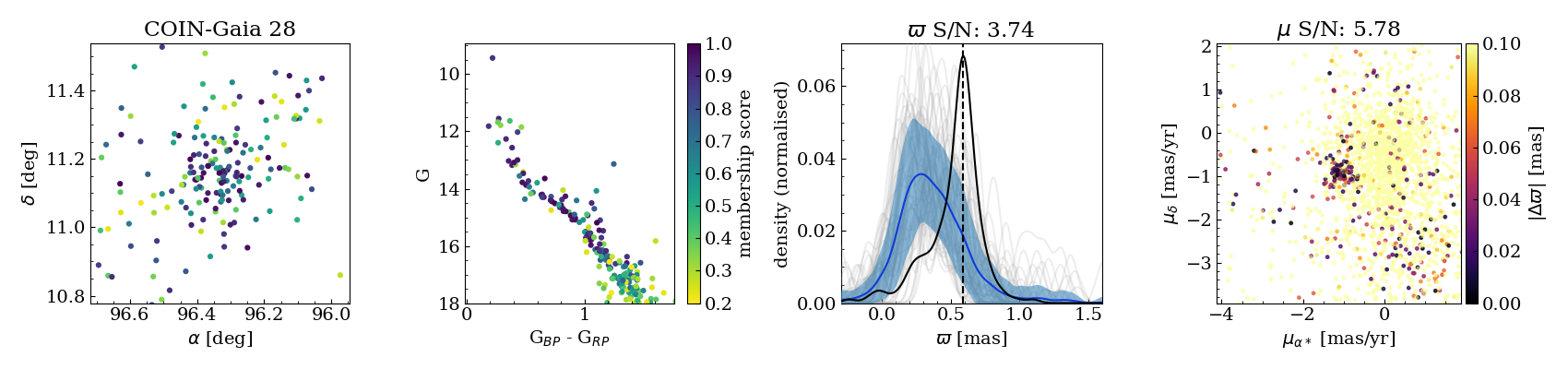}} \caption{\label{fig:example_COIN-Gaia_28} Same as Fig.~\ref{fig:example_COIN-Gaia_1} for COIN-Gaia~28.} \end{center}
\end{figure*}

\clearpage
\section{Grade B clusters}
\label{app_B}

\begin{figure*}[ht]
\begin{center} \resizebox{\hsize}{!}{\includegraphics[scale=0.5]{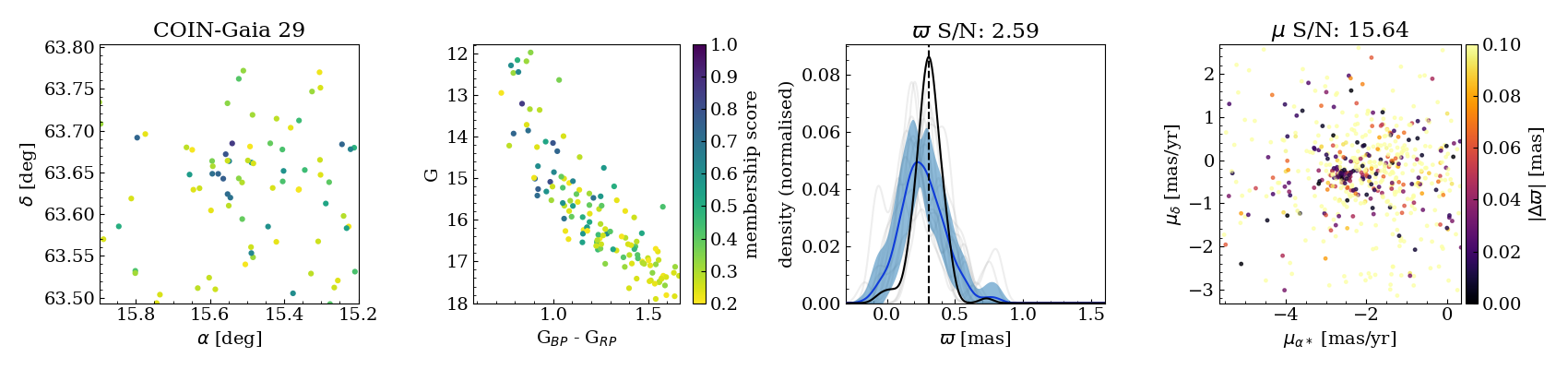}} \caption{\label{fig:example_COIN-Gaia_29} Same as Fig.~\ref{fig:example_COIN-Gaia_1} for COIN-Gaia~29.} \end{center}
\end{figure*}

\begin{figure*}[ht]
\begin{center} \resizebox{\hsize}{!}{\includegraphics[scale=0.5]{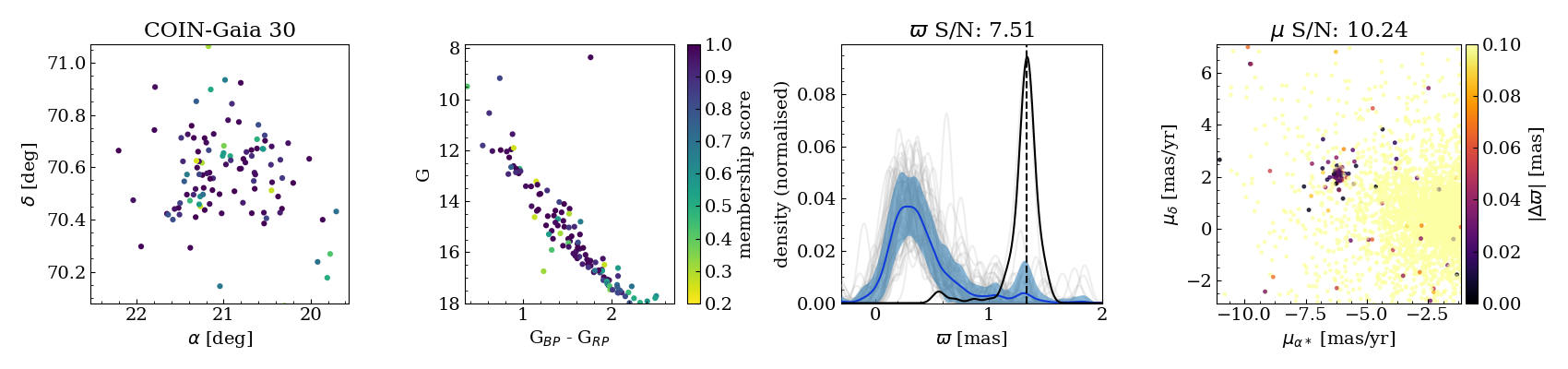}} \caption{\label{fig:example_COIN-Gaia_30} Same as Fig.~\ref{fig:example_COIN-Gaia_1} for COIN-Gaia~30.} \end{center}
\end{figure*}

\begin{figure*}[ht]
\begin{center} \resizebox{\hsize}{!}{\includegraphics[scale=0.5]{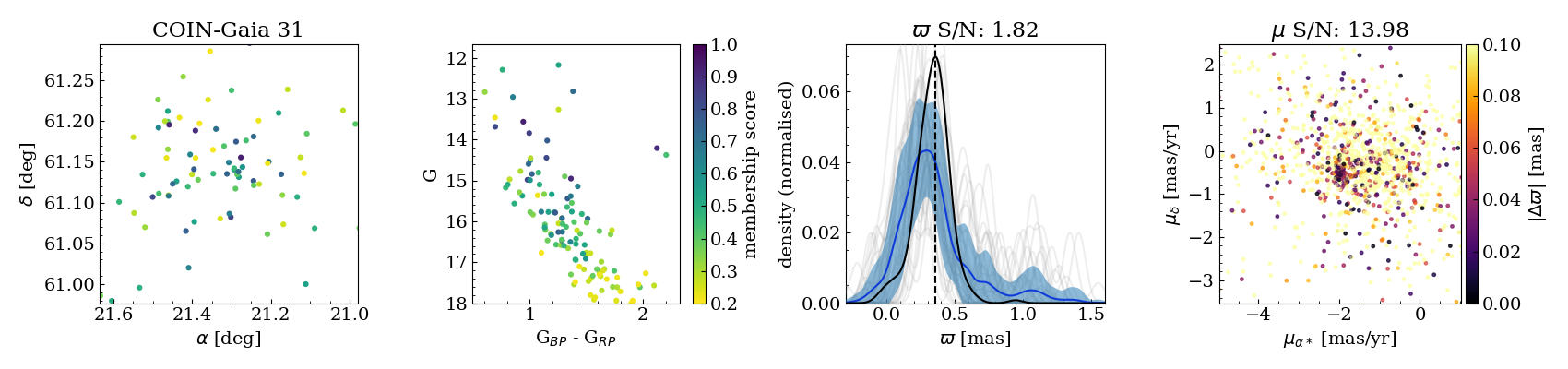}} \caption{\label{fig:example_COIN-Gaia_31} Same as Fig.~\ref{fig:example_COIN-Gaia_1} for COIN-Gaia~31.} \end{center}
\end{figure*}

\begin{figure*}[ht]
\begin{center} \resizebox{\hsize}{!}{\includegraphics[scale=0.5]{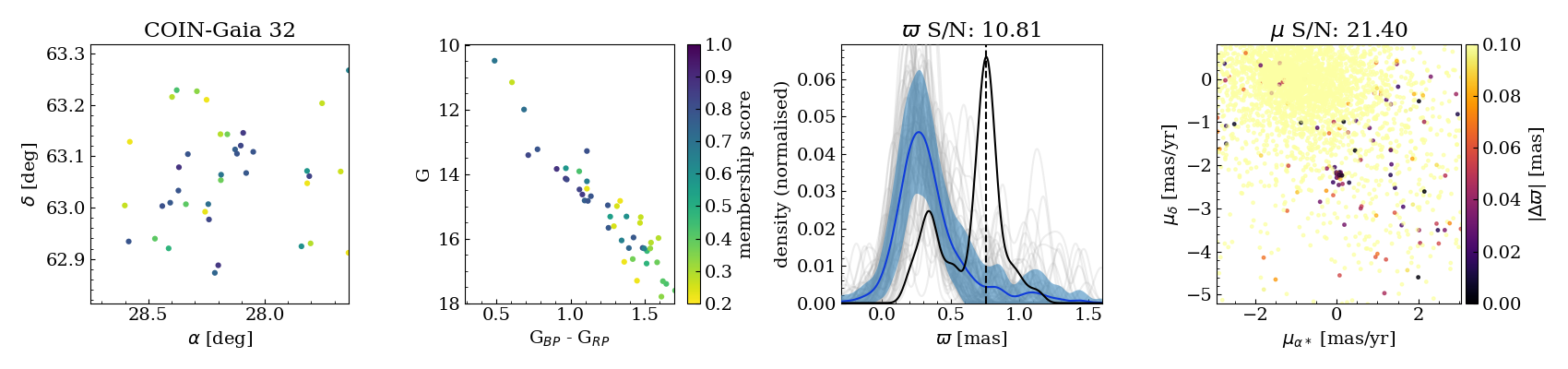}} \caption{\label{fig:example_COIN-Gaia_32} Same as Fig.~\ref{fig:example_COIN-Gaia_1} for COIN-Gaia~32.} \end{center}
\end{figure*}

\begin{figure*}[ht]
\begin{center} \resizebox{\hsize}{!}{\includegraphics[scale=0.5]{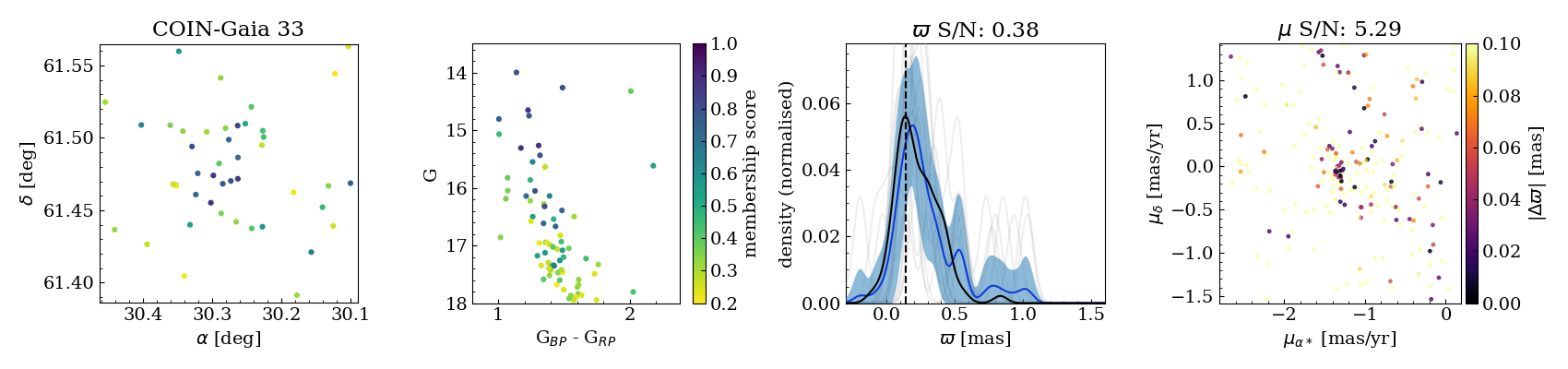}} \caption{\label{fig:example_COIN-Gaia_33} Same as Fig.~\ref{fig:example_COIN-Gaia_1} for COIN-Gaia~33.} \end{center}
\end{figure*}

\begin{figure*}[ht]
\begin{center} \resizebox{\hsize}{!}{\includegraphics[scale=0.5]{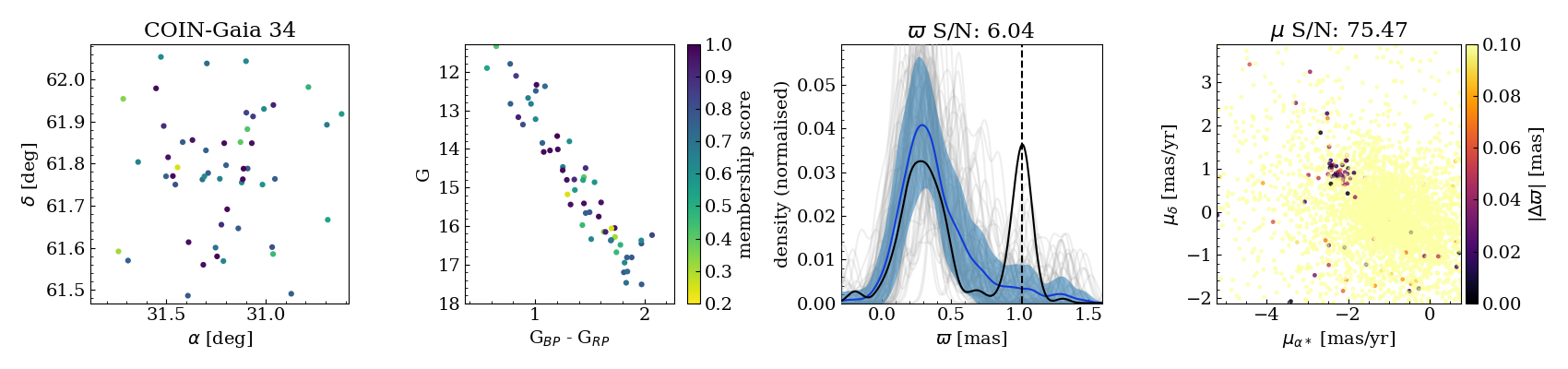}} \caption{\label{fig:example_COIN-Gaia_34} Same as Fig.~\ref{fig:example_COIN-Gaia_1} for COIN-Gaia~34.} \end{center}
\end{figure*}

\begin{figure*}[ht]
\begin{center} \resizebox{\hsize}{!}{\includegraphics[scale=0.5]{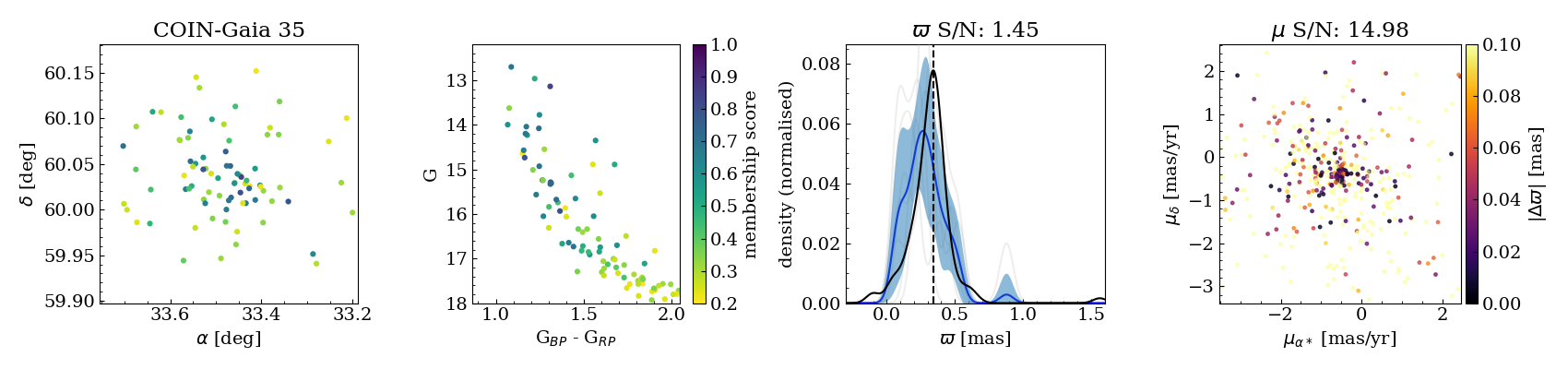}} \caption{\label{fig:example_COIN-Gaia_35} Same as Fig.~\ref{fig:example_COIN-Gaia_1} for COIN-Gaia~35.} \end{center}
\end{figure*}

\begin{figure*}[ht]
\begin{center} \resizebox{\hsize}{!}{\includegraphics[scale=0.5]{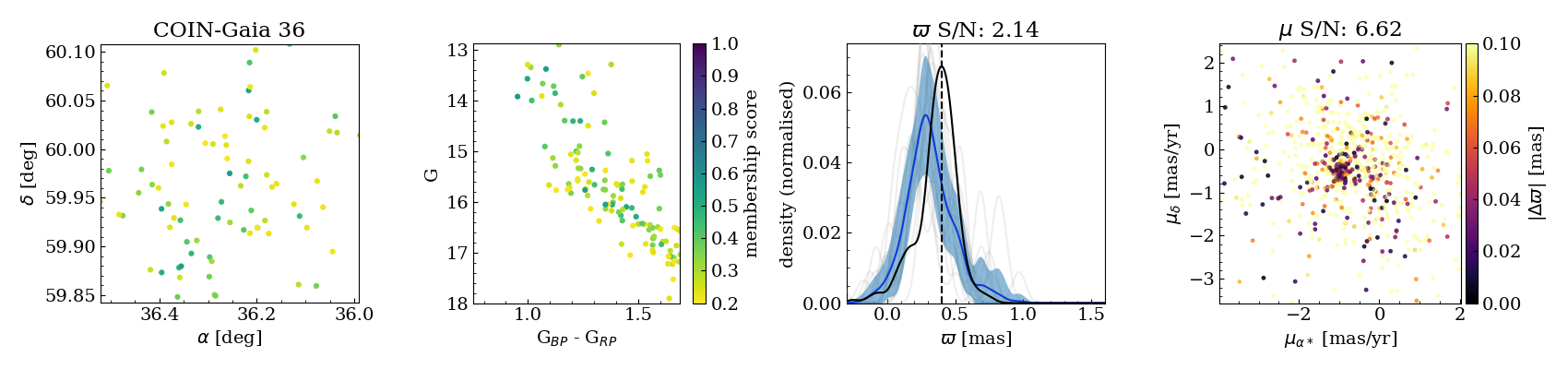}} \caption{\label{fig:example_COIN-Gaia_36} Same as Fig.~\ref{fig:example_COIN-Gaia_1} for COIN-Gaia~36.} \end{center}
\end{figure*}

\begin{figure*}[ht]
\begin{center} \resizebox{\hsize}{!}{\includegraphics[scale=0.5]{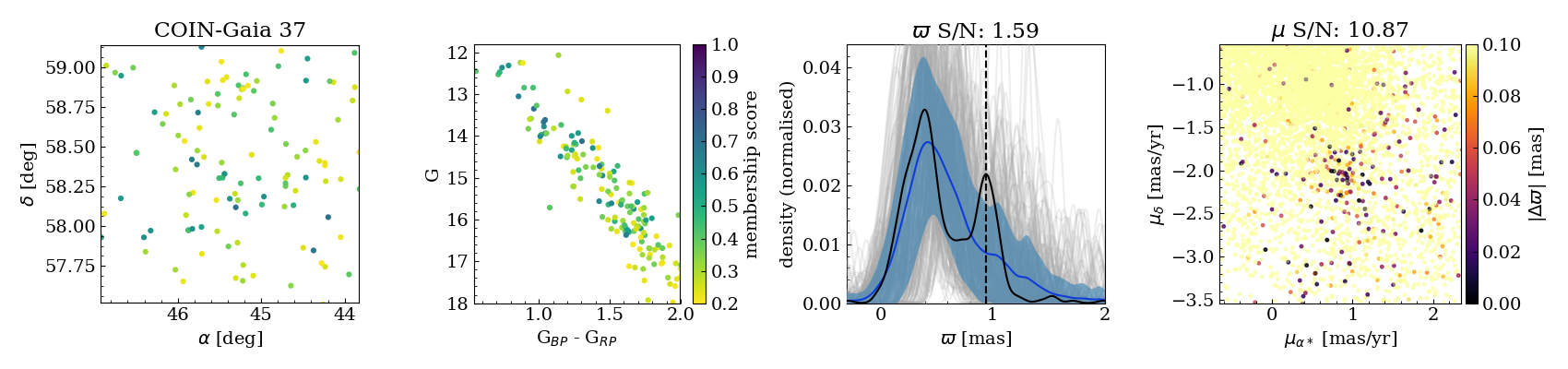}} \caption{\label{fig:example_COIN-Gaia_37} Same as Fig.~\ref{fig:example_COIN-Gaia_1} for COIN-Gaia~37.} \end{center}
\end{figure*}

\begin{figure*}[ht]
\begin{center} \resizebox{\hsize}{!}{\includegraphics[scale=0.5]{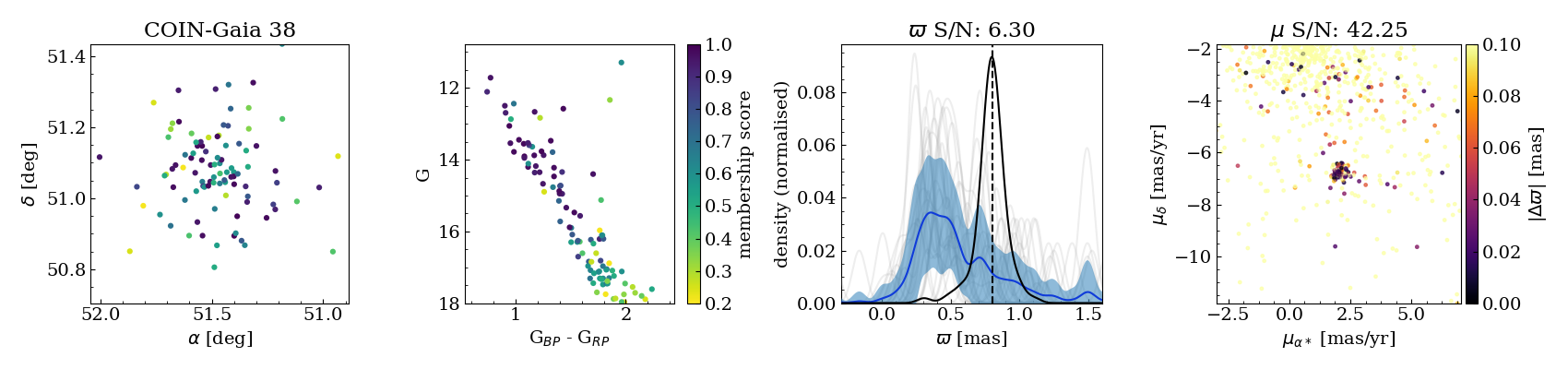}} \caption{\label{fig:example_COIN-Gaia_38} Same as Fig.~\ref{fig:example_COIN-Gaia_1} for COIN-Gaia~38.} \end{center}
\end{figure*}

\begin{figure*}[ht]
\begin{center} \resizebox{\hsize}{!}{\includegraphics[scale=0.5]{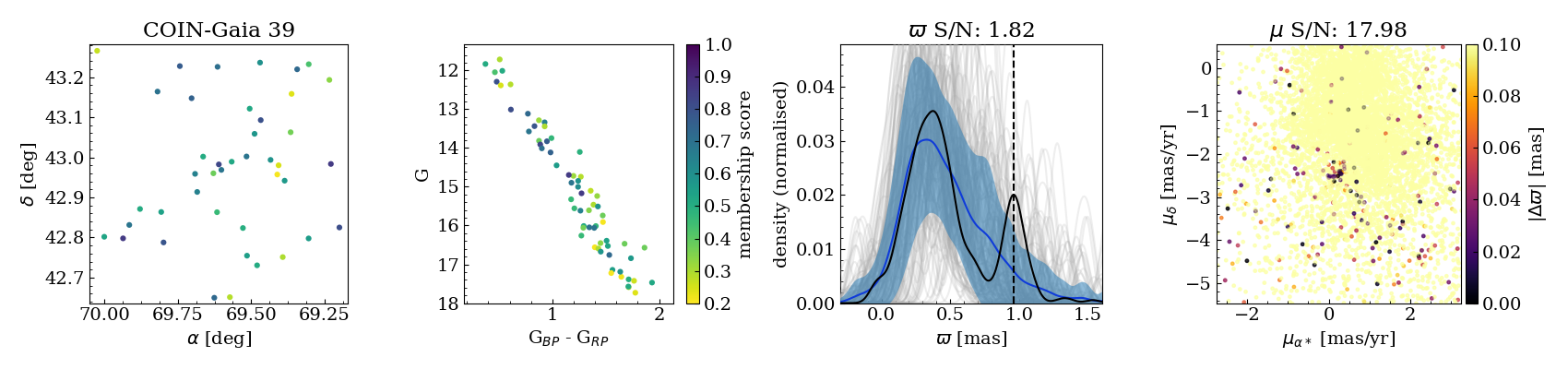}} \caption{\label{fig:example_COIN-Gaia_39} Same as Fig.~\ref{fig:example_COIN-Gaia_1} for COIN-Gaia~39.} \end{center}
\end{figure*}

\begin{figure*}[ht]
\begin{center} \resizebox{\hsize}{!}{\includegraphics[scale=0.5]{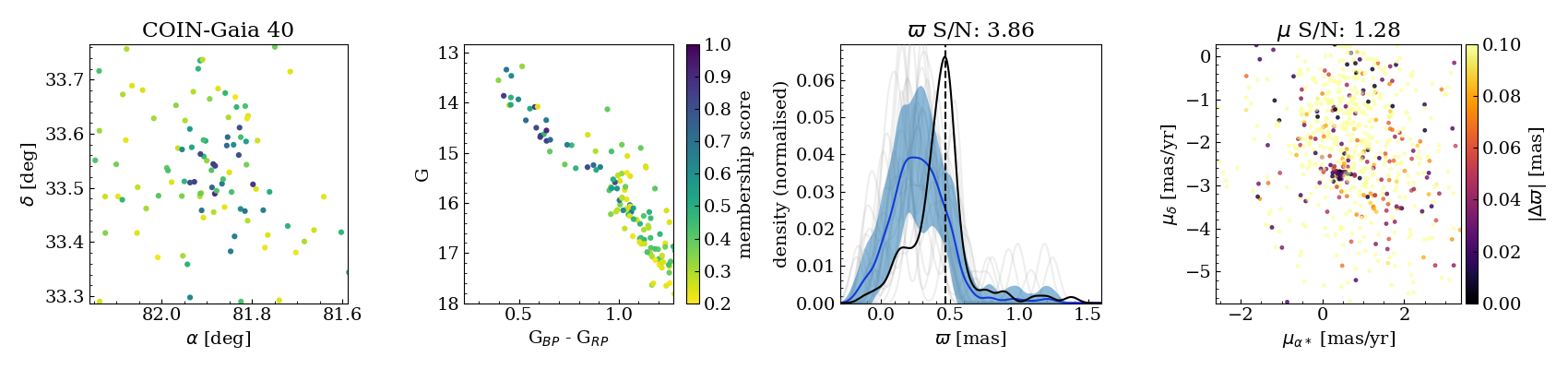}} \caption{\label{fig:example_COIN-Gaia_40} Same as Fig.~\ref{fig:example_COIN-Gaia_1} for COIN-Gaia~40.} \end{center}
\end{figure*}

\begin{figure*}[ht]
\begin{center} \resizebox{\hsize}{!}{\includegraphics[scale=0.5]{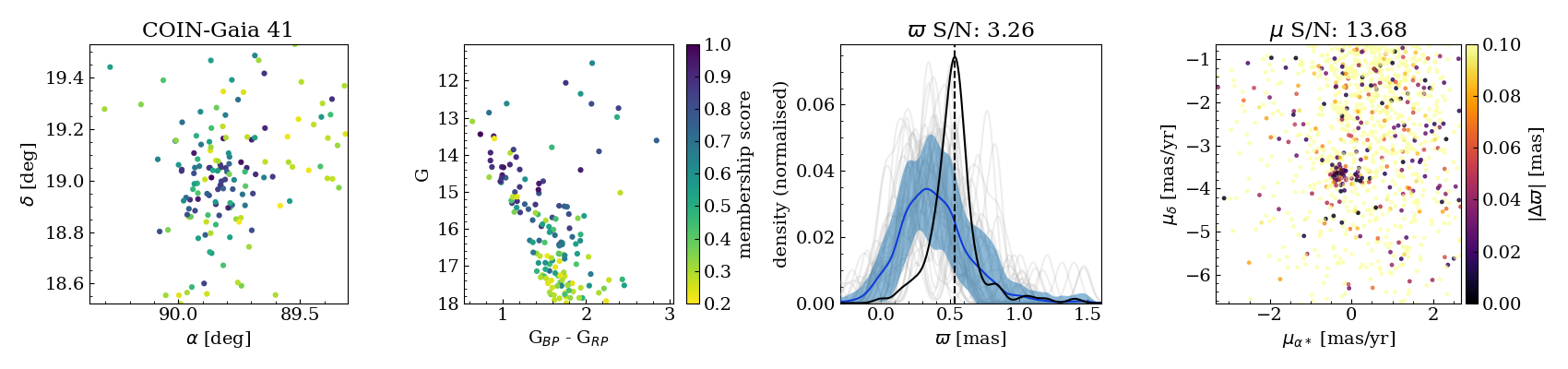}} \caption{\label{fig:example_COIN-Gaia_41} Same as Fig.~\ref{fig:example_COIN-Gaia_1} for COIN-Gaia~41.} \end{center}
\end{figure*}

\clearpage
\section{Known clusters}
\label{app_known}

\begin{figure*}[ht]
\begin{center} \resizebox{\hsize}{!}{\includegraphics[scale=0.5]{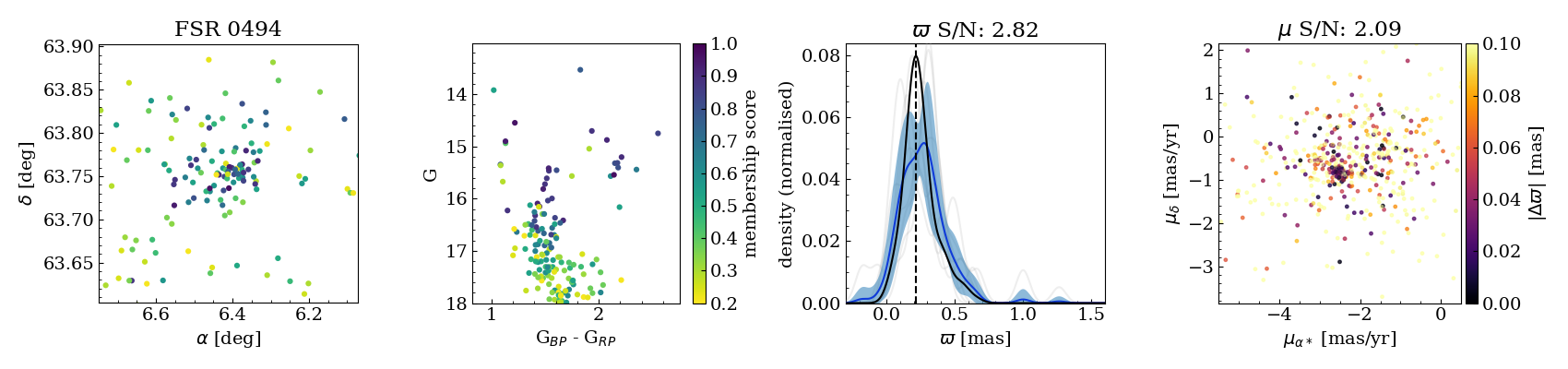}} \caption{\label{fig:example_FSR_0494} Same as Fig.~\ref{fig:example_COIN-Gaia_1} for FSR~0494.} \end{center}
\end{figure*}

\begin{figure*}[ht]
\begin{center} \resizebox{\hsize}{!}{\includegraphics[scale=0.5]{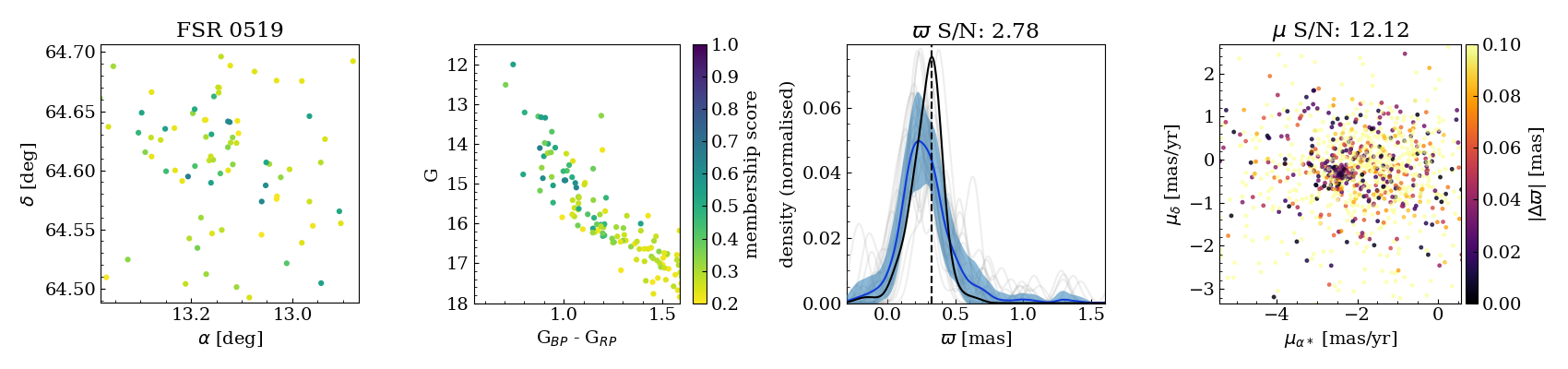}} \caption{\label{fig:example_FSR_0519} Same as Fig.~\ref{fig:example_COIN-Gaia_1} for FSR~0519.} \end{center}
\end{figure*}

\begin{figure*}[ht]
\begin{center} \resizebox{\hsize}{!}{\includegraphics[scale=0.5]{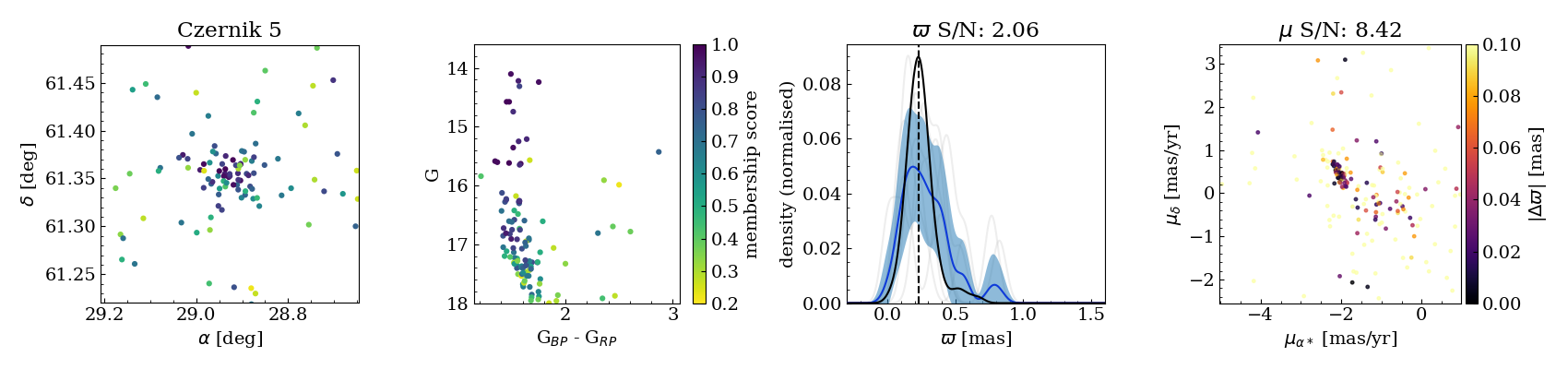}} \caption{\label{fig:example_Czernik_5} Same as Fig.~\ref{fig:example_COIN-Gaia_1} for Czernik~5.} \end{center}
\end{figure*}

\begin{figure*}[ht]
\begin{center} \resizebox{\hsize}{!}{\includegraphics[scale=0.5]{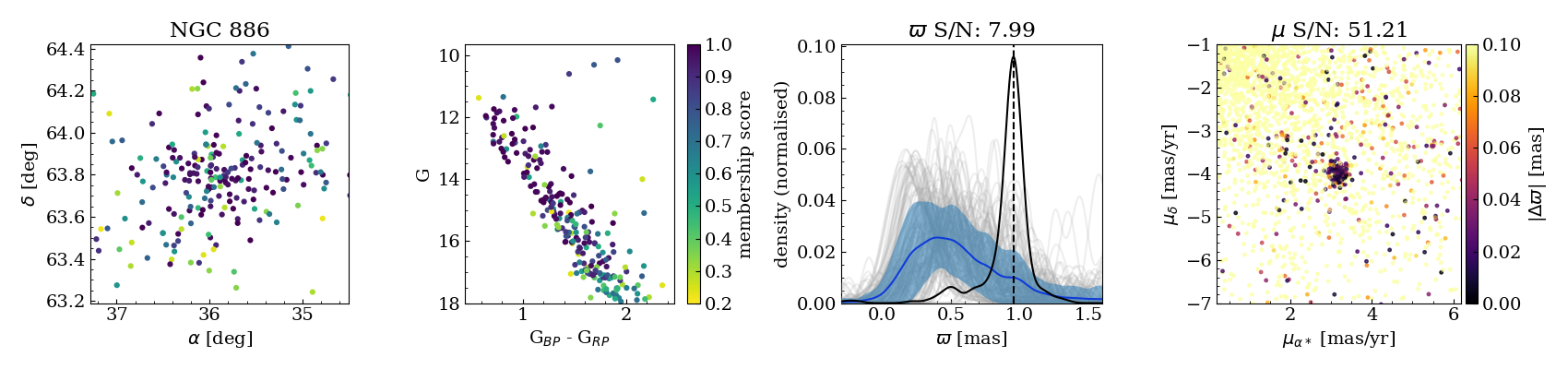}} \caption{\label{fig:example_NGC_886} Same as Fig.~\ref{fig:example_COIN-Gaia_1} for NGC~886.} \end{center}
\end{figure*}

\begin{figure*}[ht]
\begin{center} \resizebox{\hsize}{!}{\includegraphics[scale=0.5]{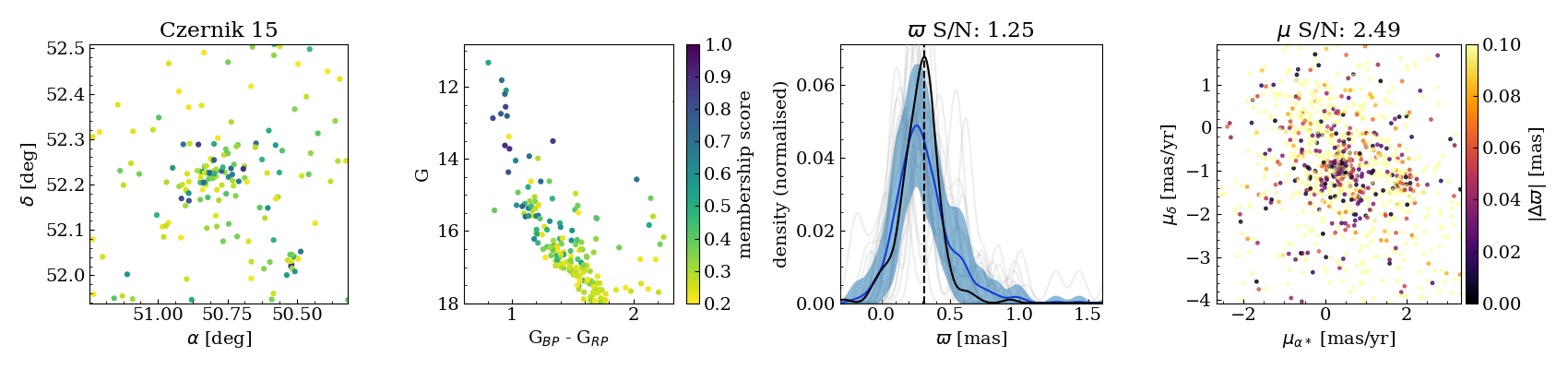}} \caption{\label{fig:example_Czernik_15} Same as Fig.~\ref{fig:example_COIN-Gaia_1} for Czernik~15.} \end{center}
\end{figure*}

\end{document}